\documentclass[aps,prb,twocolumn,superscriptaddress,showpacs,amsmath,amssymb,footinbib,longbibliography]{revtex4-2}
\usepackage[english]{babel}
\usepackage{amssymb,amsbsy,amsmath}
\usepackage{graphicx}
\usepackage[export]{adjustbox}
\usepackage{booktabs}
\usepackage{dcolumn}
\usepackage{bm}
\usepackage{subfigure}
\usepackage{color}
\usepackage{textcomp}
\usepackage[colorlinks,bookmarks=false,citecolor=blue,linkcolor=red,urlcolor=blue]{hyperref}

\newcommand{\xref}[1]{\protect\ref{#1}}
\newcommand{\figref}[1]{Fig.~\protect\ref{#1}}

\newcommand{\angstrom}{\text{\normalfont\AA}}

\begin{document}
\title{Theoretical formation of carbon nanomembranes under
  realistic conditions using classical molecular dynamics}

\author{Julian Ehrens}
\author{Florian Gayk}
\author{Patrick Vorndamme}
\affiliation{Fakult\"at f\"ur Physik, Universit\"at Bielefeld, Postfach 100131, D-33501 Bielefeld, Germany}
\author{Tjark Heitmann}
\affiliation{Fachbereich Physik, Universit\"at
  Osnabr\"uck, Barbarastr. 7, D-49076 Osnabr\"uck, Germany}
\author{Niklas Biere}
\author{Dario Anselmetti}
\author{Xianghui Zhang}
\author{Armin G\"olzh\"auser}
\author{J\"urgen Schnack}
\email{jschnack@uni-bielefeld.de}
\affiliation{Fakult\"at f\"ur Physik, Universit\"at Bielefeld, Postfach 100131, D-33501 Bielefeld, Germany}

\date{\today}

\begin{abstract}
Carbon nanomembranes made from aromatic precursor molecules are
free standing nanometer thin materials of macroscopic lateral
dimensions. Although produced in various versions for about two
decades not much is known about their internal structure. Here
we present a first systematic theoretical attempt to model the
formation, structure, and mechanical properties of carbon
nanomembranes using classical molecular dynamics simulations.
We find theoretical production scenarios under which stable
membranes form. They possess pores as experimentally
observed. Their Young's modulus, however, is systematically
larger than experimentally determined.
\end{abstract}


\maketitle

\section{Introduction}
\label{sec-1}

Many fascinating and technologically relevant carbon-based
materials, see
e.g.~\cite{GSE:APL99,Tur:AdP17,TAJ:C19,DWN:PCCP19,WEG:2DM19,Jak:B20},
cannot be simulated by quantum  
mechanical means, not even by Density Functional Theory (DFT),
since they are either too extended or structurally disordered. The latter is
for instance the case for nanometer thin carbon nanomembranes (CNMs)
of macroscopic 
lateral size, which are produced from molecular precursors
\cite{GSE:APL99,TBN:AM09,AVW:ASCN13,TuG:AM16,Tur:AdP17,DWN:PCCP19,WEG:2DM19}.
These membranes are obtained by starting with a self-assembled
monolayer (SAM) of aromatic precursors such as biphenyl,
terphenyl or longer thiols, naphthalene thiols and its longer
cousins as well as flake-like aromatic molecules grown on
e.g.\ a gold substrate \cite{AVW:ASCN13}. The SAM is
then irradiated with electrons, e.g.\ of $\sim 50$~eV with a dose
of $\sim 50$~mC/cm$^2$ \cite{MWK:AM13}; this leads to a loss of
practically all hydrogen atoms \cite{TuG:AM16}
and a cross-linking of the remaining carbon. The CNMs can be
separated from the support and used for various purposes. It is
conjectured that its properties are correlated with the
respective precursor molecules, in particular the thickness and
mechanical properties. It was demonstrated that CNMs turn into
nanocrystalline graphene when exposed to temperatures above
$500^{\circ}$~C \cite{RWT:JPCC12}. In this paper we concentrate on
theoretical investigations of CNM formation starting from the
three best-studied precursor molecules biphenyl, terphenyl and 
naphthalene thiol, see \cite{ZNA:L14} for an overview. In the past,
CNMs have only been successfully synthesized from aromatic
precursors and it was commonly believed that cyclic aliphatic
thiols would not form nanomembranes \cite{WME:JPCC12}. But this
situation changed
very recently, since it appears to be possible to use certain
non-aromatic alkanethiolate precursors too
\cite{LEK:CR05,SWA:JPCC19,Gol:PC21}. 

\begin{figure}[ht!]
	\begin{center}
		\includegraphics*[clip,width=70mm,keepaspectratio]{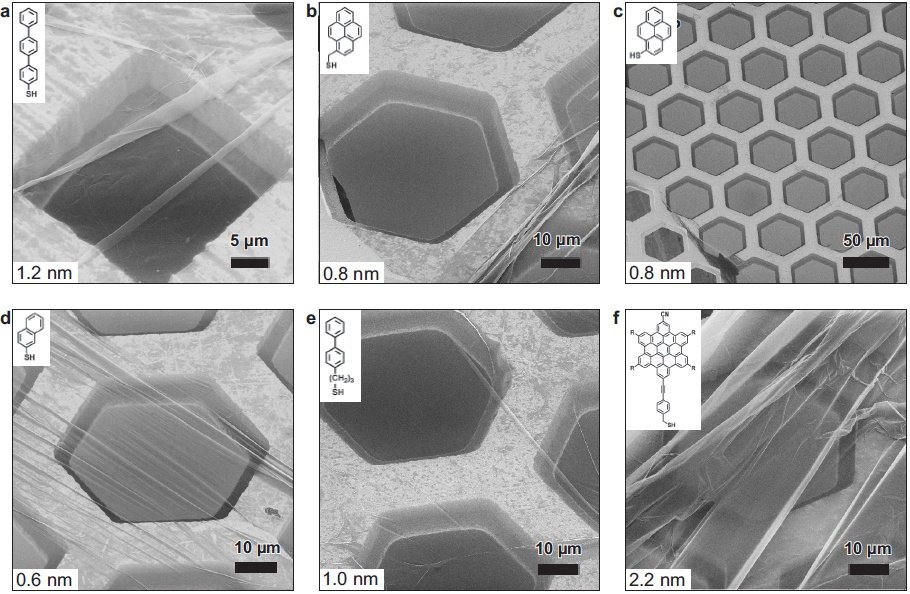}
	\end{center}
	\caption{Macroscopic CNMs
          made from various precursors resting on support
          structures. Images, made using helium ion 
          microscopy, are taken from~\cite{AVW:ASCN13} (with
          friendly permission).} 
	\label{fig:cnm-him}
\end{figure}

Although the aromatic precursor molecules biphenyl, terphenyl
and  naphthalene thiols  (BPT, TPT, NPTH, see Appendix
\ref{sams}) are 
well-characterized, not much is known about the 
internal structure of such nanomembranes \cite{MrS:BN14}. The
reason is that existing characterization methods fail to deliver
an accurate structure mainly due to the nanometer size
thickness and the tiny weight,
which, for example, does not allow accurate X-ray structure
determination or infrared spectroscopy. In addition, the
material is very likely highly disordered, which renders an
X-ray structure determination nearly impossible.

\begin{figure}[ht!]
\centering
\includegraphics*[clip,width=60mm,keepaspectratio]{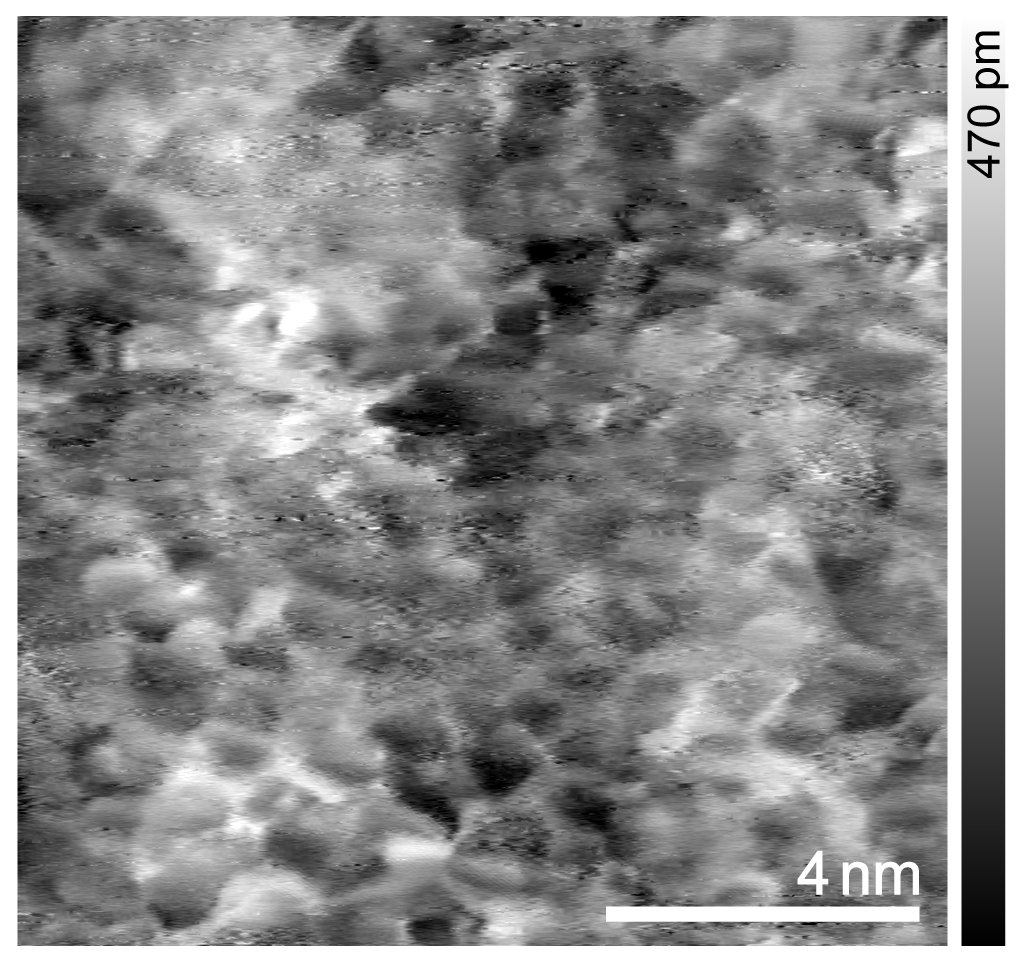}
\caption{AFM tapping mode topography image of TPT CNM on Au,
  measured at 93~K in ultra-high vacuum (amplitude set point $A = 7.6$~nm,
  center frequency $f_0 = 274.8$~kHz). The height information is
  black-to-white encoded, displaying the subnanoporous network
  of the nanomembrane.}  
\label{afm-tpt}
\end{figure}

On the other hand, the material can be produced to macroscopic
dimensions, and it is mechanically stable, see \figref{fig:cnm-him}.
Therefore, macroscopic mechanical properties, such as Young's
moduli, can be determined for such membranes \cite{ZBG:B11}. The
moduli turn out to be of the order of $10$~GPa, i.e.\ the
material is astonishingly soft compared to graphene
($1000$~GPa). It is also possible to study water
permeation \cite{yang2018rapid,YHQ:AM20} as 
well as electrical properties \cite{zhang2018large} in order to
further characterize the membranes. Investigation by means of
near edge X-ray absorption fine structure (NEXAFS) allows to
estimate the   
aromaticity, i.e.\ the amount of intact aromatic carbon rings, as
well as the $sp^2$ content still present in the CNM
\cite{TKE:L09}.  

Atomic force microscopy (AFM) delivers topographic images of
CNMs deposited on substrate material \cite{YHQ:AM20}, compare
\figref{afm-tpt}. This allows to infer information of membrane
structure on mesoscopic (nm) lateral scales, in particular the
sizes and distribution of holes and voids across the membrane.
The latter is closely related to transport properties of gases and
liquids through the membrane \cite{YHQ:AM20}.

In this article we report on first realistic and large-scale
theoretical simulations of CNMs. Before we start, we would like
to highlight the 
general problems that challenge such an investigation. This helps
to understand why such simulations have not yet been done
although CNMs already exist for about two decades.

\begin{enumerate}
\item Since quantum mechanical simulations are not at all
  feasible, we have to rely on a classical approach and thus
  unavoidably make an approximation. This holds in particular
  for the use of 
  classical carbon-carbon interactions \cite{youngsModuliOf}.
  
\item The CNM will be in a disordered metastable state, i.e.\ a
  local minimum in a huge configuration space. The true ground
  state of the material, which consists of pure carbon, would be
  a flake of graphite. It is very likely that a large number of
  disordered metastable states is actually equivalent in so far
  that they all constitute mechanically stable membranes. A
  crucial question is how much of the initial correlations
  imprinted in the precursor molecules survives and finally
  determines properties of the CNM.
  
\item How can we quantify whether a simulated structure is a
  realistic model of the true CNM? In view of the lack of
  structural information only indirect observables such as
  Young's modulus, the topographic image, or the aromaticity may
  serve as guidance.

\item In addition, the imperfectness of the CNMs, i.e.\ the
  existence and distribution of holes, that leads to the
  fascinating property of water filtration
  \cite{yang2018rapid,YHQ:AM20}, can also serve as a clue. To
  this end larger CNMs have to be simulated in order to minimize
  finite-size and boundary effects.
  
\end{enumerate}

For our simulations we employ classical molecular dynamics as
implemented in the publicly available large atomic/molecular
massively parallel simulator (LAMMPS) \cite{Pli:JCP95}.
Our previous studies have shown that the potentials and algorithms
implemented in LAMMPS are accurate to a large extent for other
carbon-based systems as e.g.\ diamond, graphene, or nanotubes
\cite{youngsModuliOf}. The environment-dependent interatomic
potential (EDIP) of Marks \cite{Mar:PRB00,MTS:PRL19},
not implemented in LAMMPS, appears superior in several contexts,
and is thus also employed \cite{TSM:C16,MTS:PRL19,TAJ:C19}.

In order to incorporate at least the gross features of the
production process we decided to mimic the formation of the CNM
as a dynamical process that consists of excitation, compression
as well as expansion and equilibration. This goes far beyond the
more quasistatic approach of Ref.~\cite{MrS:BN14} used earlier.

We can summarize our findings as follows: It is indeed possible
to simulate mechanically stable CNMs. We find production
scenarios under which the membranes possess holes of the correct
size as determined by AFM measurements \cite{YHQ:AM20}. We can also
theoretically determine the Young's modulus. It is systematically
larger than experimentally found \cite{ZBG:B11}. The reasons will be
discussed later in the paper. We can also relate the number of
perfect hexagons in the classical structures to the
experimentally deduced aromaticity. It turns out that both,
experiment and theory suggest, that stable CNMs contain a
drastically reduced number of aromatic rings compared to their
precursors. The broken-up rings seem to be a necessary
prerequisite to deliver the ``glue'' for the stabilization of
the membrane.

The article is organized as follows. In Sec.~\xref{sec-2} we
shortly repeat the essentials of classical molecular 
dynamics calculations as well as the technical details employed
for the simulations. The main section~\xref{sec-3} is devoted
to the simulations of CNMs as well as to the 
determination of their physical properties. In the appendix we
present first attempts to generate AFM images related to our
simulations. 
The article closes with discussion and conclusions.

\section{Method and technical details}
\label{sec-2}

\subsection{Classical carbon-carbon interaction}
\label{sec-2-1}

A realistic classical carbon-carbon interaction must be able to
account for the various $sp^n$--binding modes. The program
package LAMMPS offers several of such
potentials, among them those developed by Tersoff and Brenner in
various versions \cite{Ter:PRB88,Bre:PRB90,BSH:JPCM02} as well
as new extensions built on the original potentials.

In addition to the implemented potentials we are going to use
the improved EDIP potential by 
Marks \cite{Mar:PRB00,MTS:PRL19} for some of our simulations,
which so far 
is not included in standard versions of LAMMPS. Taking this
potential as an example, we want 
to qualitatively explain how such potentials work. These
potentials comprise density-dependent two- and three-body
potentials, $U_2$ and $U_3$ in this example respectively,  
\begin{eqnarray}
\label{E-2-1}
U\left(\vec{R}_1,\dots,\vec{R}_N\right) 
&=& \sum_{i=1}^{N} 
\Big\{
\sum_{\substack{j=1\\j\neq i}}^{N} U_2(R_{ij}, Z(i)) 
\\
&+&
\sum_{\substack{j=1\\j\neq i}}^{N} \sum_{\substack{k=j+1\\k\neq
    i}}^{N} U_3(R_{ij},R_{ik},\theta(i,j,k),
Z(i))\Big\}
\nonumber
\end{eqnarray}
which account for the various binding modes. This is achieved by
an advanced parameterization in terms of a smooth coordination
variable $Z(i)$ as well as by appropriate angle dependencies
$\theta(i,j,k)$. The EDIP potential employs a cutoff of
3.2~\AA\ and a dihedral penalty.

Another popular option for carbon-carbon (C-C) interactions is
AIREBO \cite{airebo}, which also includes the necessary
implementation for carbon-hydrogen (C-H) interactions. This
potential is employed in our simulations when the virial per
atom is needed, since this is not yet implemented for
EDIP.

\subsection{Modeling of the membrane}

\begin{figure}[ht!]
	\begin{center}
		\includegraphics*[clip,width=65mm,keepaspectratio]{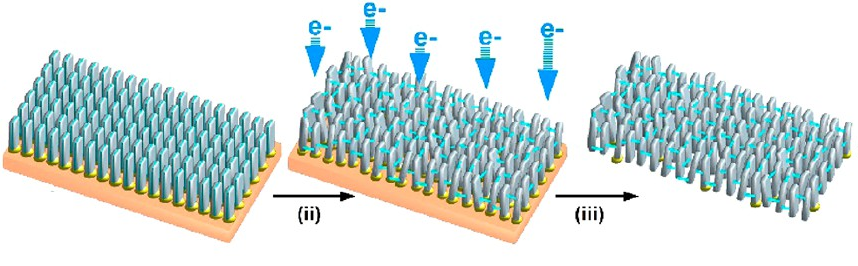}

		\includegraphics*[clip,width=80mm,keepaspectratio]{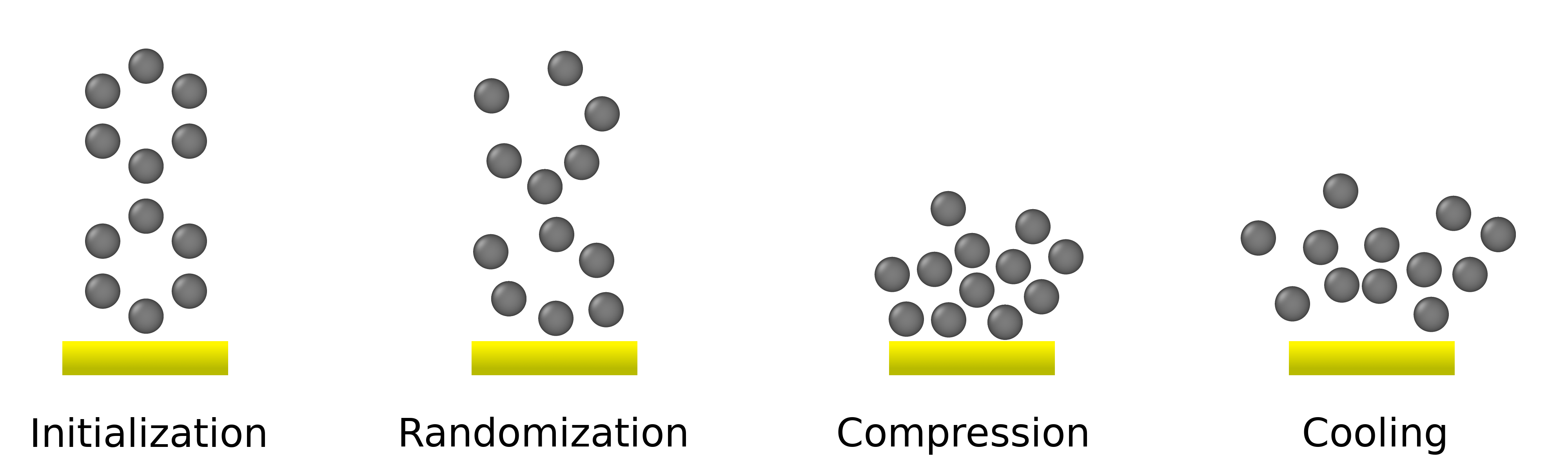}
	\end{center}
	\caption{Top: Sketch of the experimental synthesis of a
          CNM, taken from ~\cite{AVW:ASCN13}. 
          Bottom: Sketch of the theoretical four step synthesis
          model of a CNM starting from a precursor corresponding
          to BPT.} 
	\label{fig:synth-scheme}
\end{figure}

Modeling of a membrane is achieved through the
following steps inspired by the experimental procedure as
depicted in \figref{fig:synth-scheme} and explained in detail in
\cite{AVW:ASCN13}. Our simulations include only carbon
atoms, since the precursor molecules of the SAM lose
practically all hydrogen atoms during electron irradiation, see
experimental verification in \cite{TKE:L09,TuG:AM16}, such that the
remaining carbon skeletons interlink to form the final CNM
which thus is a form of pure carbon. All other atoms such as
sulfur of the thiol are also neglected right from the beginning.
The loss of carbon is small during the production process
\cite{AVW:ASCN13,TuG:AM16}, Table~\xref{tab:xps},
but will nevertheless be addressed by us since
it might have an impact on the formation of holes.

\begin{enumerate}
\item Formation of a self-assembled monolayer (SAM) from a
  selection of various precursor molecules on a gold substrate
  is initiated by placing carbon atoms above a gold surface at
  positions they would have in the respective precursor
  molecules (Initialization).
  
  Also, we replace the computationally
  expensive array of gold atoms representing 
  the substrate by a repulsive Lennard-Jones wall potential  
\begin{align}\label{eqn:lj}
	V(r) = 4 \varepsilon \left[
          \left(\frac{\sigma}{r}\right)^{12}
          - \left(\frac{\sigma}{r}\right)^6 \right],
\end{align}
with its minimum $r_{min} = \sqrt[6]{2} \cdot\sigma$ at the
bottom of the simulation box $z_{\text{lo}}$ and parameters
for the C-Au interaction 
$\varepsilon_{C-Au} \approx 0.29256~\text{kcal/mol}\approx
0.012695$~eV and $\sigma_{C-Au} \approx 2.99$~\angstrom\ taken
from~\cite{onCorrelationEffect}. It 
should be mentioned however that this also leaves us with no
structure of the substrate (except for the structural parameters
of gold taken for the initial placement of the precursor molecules),
which could have some influence on the formation process.

\item   Then, specific starting conditions are imposed by 
  tilting or randomly moving some or all molecules and by either
  removing some of the atoms or whole molecules to e.g.\ mimic
  defects in the experimental process (Randomization), see
  \cite{TSM:C16,SMM:PRL18,MTS:PRL19} 
  and Table~\ref{tab:xps}.

\item Experimentally, after low-energy electron irradiation of
  the SAM, crosslinking of the molecules induces the formation
  of the CNM. Theoretically, the electron irradiation is modeled by a
  vertical force gradient being applied to the atoms; it is
  linear and decreasing with height (Compression). It is assumed
  that secondary electrons actually cause most of the
  bond-breaking and crosslinking. The effect of secondary
  electrons is e.g.\ modeled by lateral forces on randomly
  selected molecules. In reality, processes probably follow a
  sequential order on a short time scale. This is neglected in
  the present simulations, in particular since it is not clear whether
  and how such correlations survive in the course of relaxation
  towards the final structure which happens on much longer time
  scales. 

\item The model system is then allowed to relax towards an
  equilibrium structure according to a thermostat dynamics
  (Nos\'e-Hoover or Langevin) with a temperature that decreases
  linearly in time (Cooling). This corresponds to the fact that
  the gold support also acts as a very efficient heat sink
  during the synthesis process. 

\end{enumerate}

Measurements by X-ray photoelectron spectroscopy (XPS) presented
in Table~\ref{tab:xps} provide 
a qualitative measure for the modeled membranes. The thickness of
the membrane should remain close to the thickness of the
original SAM, since there is only little loss of carbon during
irradiation \cite{AVW:ASCN13,TuG:AM16}. 

\begin{table}[ht!]
\centering
\caption{Experimental thickness and carbon loss as determined by
  XPS measurements~\cite{AVW:ASCN13}.}
\label{tab:xps}
\begin{tabular}{l|r|r|r}

     & SAM thickness & CNM thickness & C loss \\ 
\hline
	 BPT  & {10}{\AA}      & {9}{\AA}       & 5 \% \\
	 TPT  & {13}{\AA}      & {12}{\AA}      & 4 \% \\
	 NPTH & {6}{\AA}       & {6}{\AA}       & 9 \% \\\hline
\end{tabular}
\end{table}

We divide the outcomes of our simulation procedure into four
categories depending on the parameters used: 1) weak
randomization, i.e.\ there is only some randomization of atom
coordinates, 2) randomization and compression, i.e.\ after
randomization a vertical force is applied, 3) randomization,
compression and lateral force that acts on some selected
molecules, and 4) randomization, compression and randomly
excluding molecules from the simulation.

\subsection{Determination of the Young's modulus}
\label{sec-2-2}

With the structures generated, we choose the Young's modulus as
our observable of choice as it allows comparison with
experimental results (note that electronic properties cannot be
covered by classical molecular dynamics). These calculations are
realized in two ways:
\begin{enumerate}
    \item We adapt LAMMPS' own ELASTIC code as available in the
      example repository \cite{lammpsexamples} to our
      needs which derives the Young's modulus from the curvature of
      the potential Energy $U$. For this we use our own
      implementation of the EDIP.
    \item We use a  dynamical approach that stretches the membrane
      (stress vs. strain), thereby allowing for deformation and defect
      formation, and derives the modulus from the linear region of
      the stress-strain curve. Due to the lack of the virial per
      atom in our implementation of EDIP, we use the AIREBO
      potential for this type of simulation. 
\end{enumerate}

The Young's modulus $E$ in 
the ground state, i.e.\ at temperature $T=0$~K, 
can be evaluated from the curvature of $U$ at the ground state
configuration (the kinetic energy is zero) \cite{HGB:PRL98}
\begin{equation}
\label{E-2-2}
E_V = \frac{1}{V_0} \left(\frac{\partial^2 U}{\partial \alpha^2}\right)_{\alpha=1}
\ ,
\end{equation}
where $\alpha$ is the factor by which all
positions are scaled along the direction of the dimensionless
unit vector $\vec{e}_{\alpha}$, 
i.e.
\begin{equation}
\label{E-2-2b}
\vec{x}_i \rightarrow \vec{x}_i + (\alpha - 1) \
\vec{e}_{\alpha}\cdot\vec{x}_i \ \vec{e}_{\alpha}
\ .
\end{equation}
$V_0$ denotes the cuboidic volume of the sample in equilibrium. 

There is however a challenge when it comes to the definition of
volume of a CNM due to its irregular internal
structure. Thus one has to find ways to approximate the volume,
which introduces inherent uncertainty into the results, since
the variation of thickness is of the same order as the thickness
itself. Possible approaches are presented in Sec.~\ref{postpro}. 

Another approach is to derive the Young's modulus from the
relationship between stress $\sigma$ and strain $\varepsilon$ in
the linear part of a stress-strain-curve as employed by
materials science for macroscopic materials, i.e.\ by determining 
\begin{align}
  E = \frac{\Delta \sigma}{\Delta \varepsilon}
  \ ,
\end{align}
which can be done in classical molecular dynamics by moving
clamped parts of the material similar to experiments for
material characterization. This is 
not directly transferable to real CNMs as they cannot
be investigated this way due to their restricting size. The
alternative way to experimentally characterize such thin
membranes is by 
performing a bulge test \cite{ZBG:B11} where the
deflection of a membrane under pressure is measured by the tip
of an atomic force microscope. This has been modeled as a
molecular dynamics simulation for graphene in \cite{jun2011size}.
Here we do not use this method since there is no well defined
profile of curvature of the membrane while for graphene there is
even a formula for expressing the maximum height of the graphene
sheet depending on the applied pressure difference. Also, one
might have to resort to bigger molecules for the gas (for
graphene hydrogen is used) when this model is transferred to
CNMs as the holes possibly allow for gas molecules to pass
through the membrane, making it hard to keep track of applied
pressure when there is a vacuum on the other side.

\subsection{Simulation setup}
\subsubsection{General setup}

All simulations are done with shrink-wrapped boundary conditions
(\verb|boundary s s s|), which causes the simulation box to be
non-periodic. We use \verb|metal| units and a time step of
$0.0001$. The primary \verb|pair_style|, i.e.\ potential we
employ is \verb|airebo| in the current parameterization of LAMMPS
stable release from 22. August 2018 in \verb|CH.airebo| with a
\verb|cutoff| of $3.0$ as well as Lennard-Jones and torsion
flags being set to $1$. For the EDIP we use our own
implementation of Nigel Marks' carbon-carbon
potential \cite{gayk} that is not available in the official LAMMPS
repositories. We perform constant NVE integrations as
implemented in LAMMPS, i.e.\ integrations with conctant particle
number, constant volume, and constant energy.

\subsubsection{Modeling process details}

As for specific modeling setups we make use of
\verb|velocity create| to introduce randomization and
\verb|fix addforce| to apply downward momentum transfer to
either all or 
some molecules or atoms. Selection of atoms and molecules is
done by \verb|group|ing the desired amount and preprocessing of
the LAMMPS input script by external scripting means.  
The bottom wall representing the gold substrate is achieved by a
repulsive Lennard-Jones potential as described above in
(\ref{eqn:lj}) using \verb|wall/lj126|. Wall potentials on the
lateral sides of the membrane that prevent unwanted spreading
are of the type \verb|wall/harmonic| with a \verb|cutoff| large
enough to prevent atoms from passing the wall within a time step.  

After the application of force or thermal randomization of atoms
and molecules, the CNM is highly excited and has
thus to be cooled. This is achieved by means of a Langevin-type
thermostat as implemented in LAMMPS. The final structure of the
membrane depends strongly on the parameters used for the 
thermostat, i.e.\ the damping rate of the Langevin thermostat has
direct influence on how much time the atoms have to spread in
the $z$-direction and thus making the membrane thicker and less
dense. This has measurable influence on the Young's modulus and
is thus kept at the recommended best practice damping rate
related to the time step, which in this case gives a damping of
$0.001$. Other thermostats can be tweaked such that results are
very similar to each other. 

\subsubsection{Postprocessing}\label{postpro}

Postprocessing takes care of adjusting the Young's moduli with
respect to the volume of the system. However, as this is not
well-defined one has several options to calculate the
volume. The first and simplest is to take the size of the
simulation box, which does not account for voids. The second more involved
method is to create a surface volume of the CNM
that tries to minimize superficial empty volumes thus creating a
shrinkwrap-like representation of the membrane's volume. The
latter and all of the visualization tasks are done with the
software Ovito \cite{Stu:MSMSE10}.

\section{Theoretical Investigations}
\label{sec-3}

\subsection{Weak randomization}
\label{sec-3-1}

\begin{figure}[ht!]
\centering
\includegraphics*[clip,width=65mm,keepaspectratio]{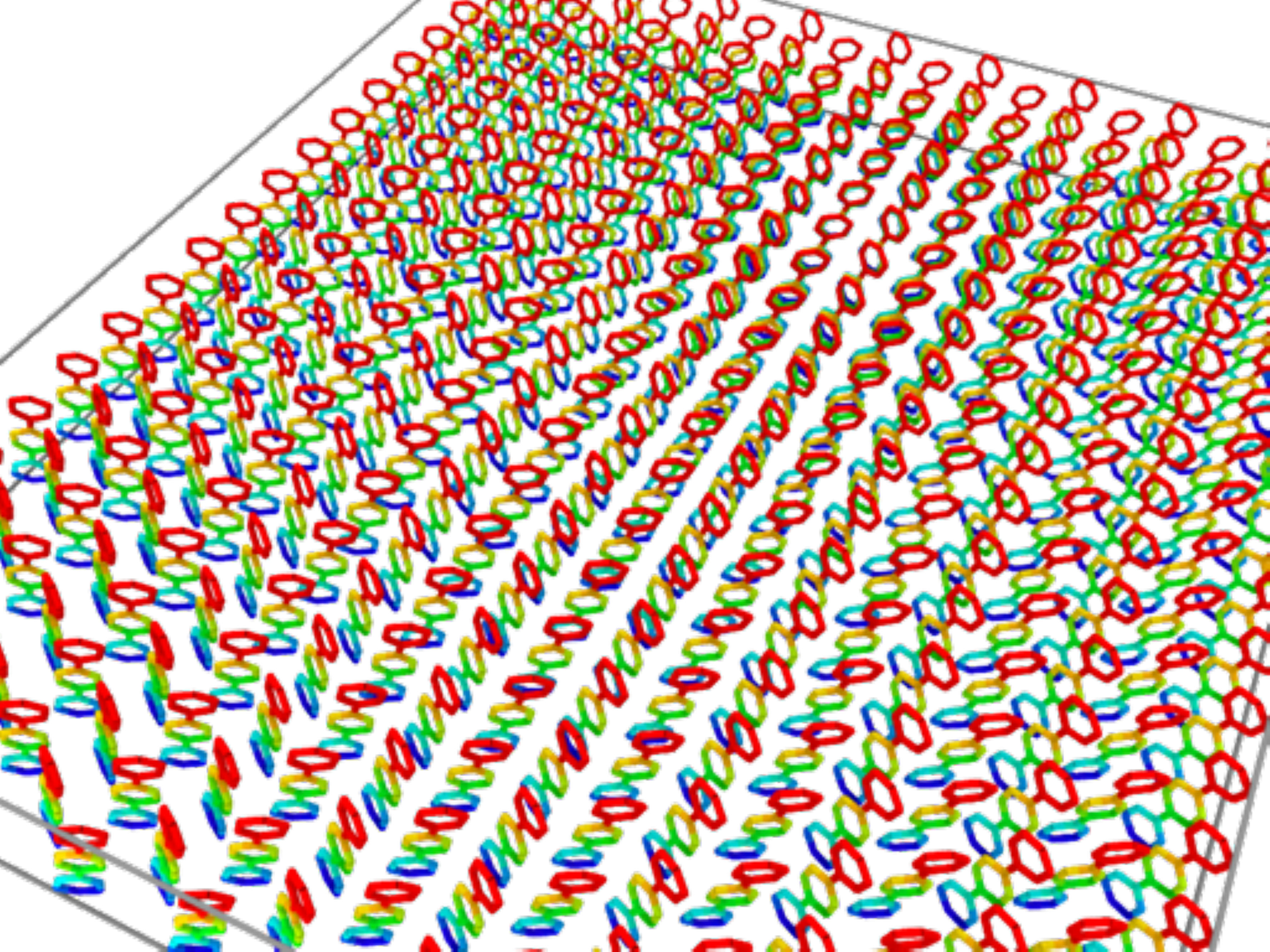}
\caption{A typical result of minimization of a weakly randomized TPT-based
  CNM. Color represents position along
  $z$-direction (blue to red as on regular maps). The
  $x$-coordinate runs along the visible rows from bottom to top
and the $y$-coordinate perpendicular to that, i.e.\ from left to
right.} 
\label{wr-b}
\end{figure}

In this section we will discuss the results of the molecular
dynamics simulations where only weak randomization is employed.
As a reminder, randomization is one theoretical means to model
the excitation of the SAM due to the electron bombardment.
To this end, atoms of the SAM are given random displacements
with respect to the initial configuration. The isotropic
randomization corresponds to a temperature in the range
$300-1100$~K. The system is then cooled down to find a stable
configuration. 

Resulting nanomembranes, as shown for TPT in \figref{wr-b}, mostly
retain the initial structure of the SAM. The excitation energies
are not sufficient to break up carbon bonds and to create a
disordered structure. This can be clearly seen in \figref{wr-b}
where in addition the color code represents the height above the
gold support. The membrane consists of weakly interacting
upright terphenyls. Similar results have been obtained in the quasi
static approach of Ref.~\cite{MrS:BN14}.
Nevertheless, the membrane represents a
bound state due to the attractive long range component of the
carbon-carbon interaction, and it is mechanically
stable. However, we consider 
such membranes unrealistic since they do not form any
non-regularities such as holes observed by AFM
and obviously needed for water permeation \cite{YHQ:AM20}.
Membranes of all three investigated precursor molecules, BPT,
TPT, and naphthalene, behave in the same way; we therefore show
only a case of TPT.

The high density of the membrane is also reflected in
the Young's moduli presented in Table~\ref{tab:wr}. Here and in
the 
following, we show two values for each Young's modulus to
address the problem of volume in the definition of the Young's
modulus as previously discussed -- the first adjusted to the
volume of the simulation box and the second to the
shrink-wrapped surface volume. Due to the very anisotropic
structure of the simulated CNMs also the moduli are very
anisotropic. 
The rather small Young's modulus in the $y$-direction
perpendicular to the rows for all
precursor molecules can be explained by visual inspection of the
carbon-carbon bonds existent in the membrane, compare
\figref{wr-b}. Due to the nature of the self-assembled
monolayer, there is a larger distance between 
neighboring rows of molecules in the $y$-direction than there is
between molecules in the $x$-direction. This is also the reason
for stronger bonds forming in the $x$-direction.

\begin{table}[ht!]
	\centering
	\caption{Young's moduli in $x$- and $y$-direction
          adjusted to volume of simulation box$|$surface
          volume for single realizations of membranes.  
          For direction of coordinates compare
          \figref{wr-b}. Index 1 denotes method 1 (EDIP and 
          curvature), index 2 denotes method 2 (AIREBO and stress).}
	\begin{tabular}{c|c|c|c|c}
        & $E_{x,1}$/GPa & $E_{x,2}$/GPa & $E_{y,1}$/GPa & $E_{y,2}$/GPa \\\hline
		BPT & $266|576$ & $33|71$ & $139|301$ & $66|143$ \\
		TPT & $373|398$ & $52|56$ & $40|43$ & $287|309$ \\
        NPTH & $576|734$ & $124|158$ & $77|98$ & $22|28$ \\\hline
	\end{tabular}
	\label{tab:wr}
\end{table}

\subsection{Randomization and compression}
\label{sec-3-2}

\subsubsection{Vertical momentum only}

\begin{figure}[ht!]
\centering
\includegraphics*[clip,width=65mm,keepaspectratio]{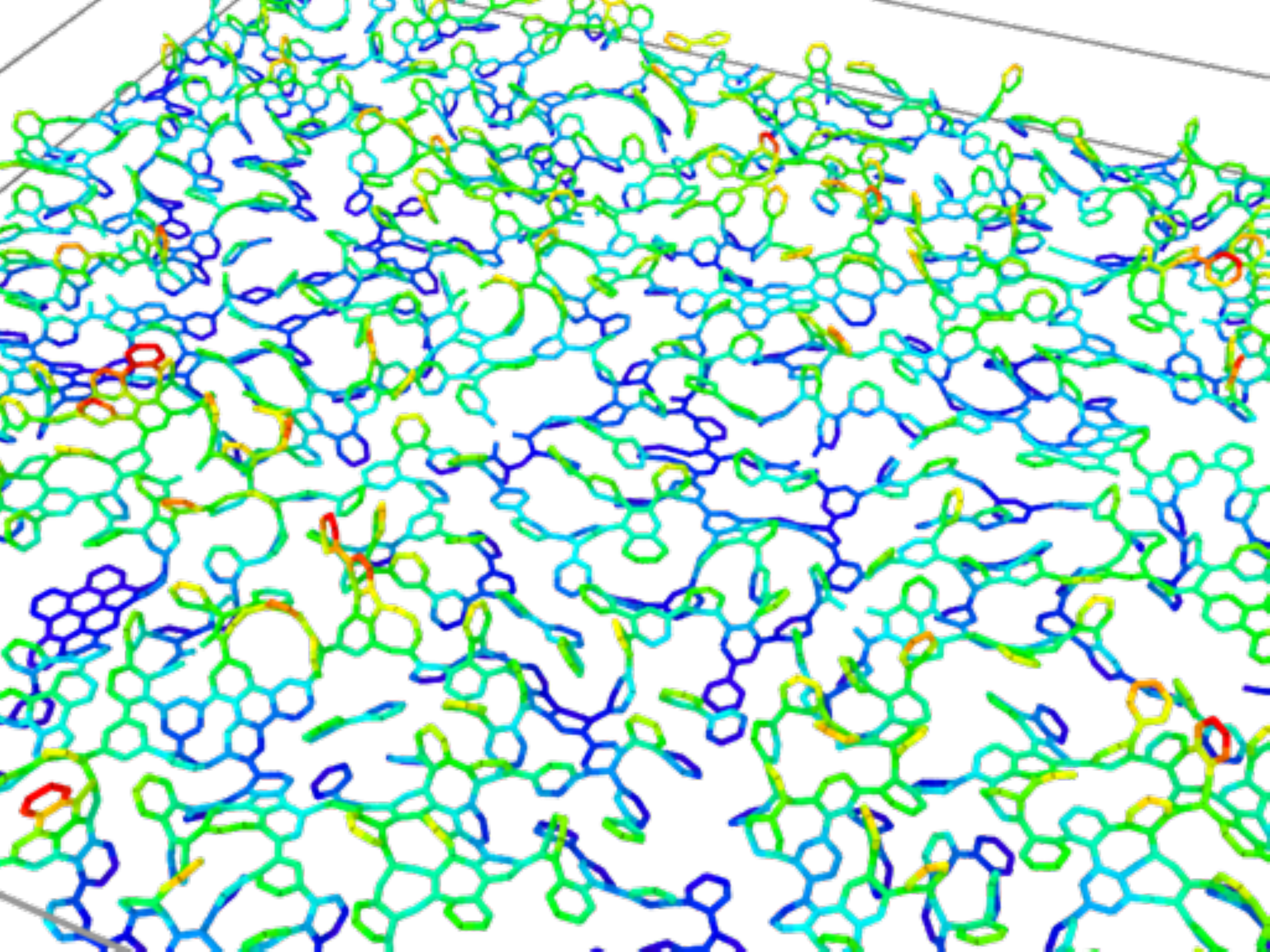}
\caption{A typical result of vertical momentum dynamics applied
  to BPT-SAM, $T=700$~K, $k=60~\frac{\text{eV}}{\angstrom}$. Color represents position along
  $z$-direction (blue to red as on regular maps).
  $x$- and $y$-directions as in \figref{wr-b}.}  
\label{v-b}
\end{figure}

A more realistic approach is to apply a vertical momentum to the
molecules of the self-assembled monolayer in direction of the
gold substrate to simulate the momentum transfer of electrons to
the atoms. Since most of the electrons' energy should be
absorbed at the top of molecules, we use a linear profile for
the applied force using the LAMMPS command \verb|addforce|,
i.e.\ $F = -k \cdot (z-z_{\text{lo}})$, where $z_{\text{lo}}$ is
the $z$-coordinate of the gold surface. During the time
evolution of this procedure, atoms 
will be compressed towards and reflected away from the
substrate. Time evolution is stopped when the height of the
membrane approaches the initial monolayer height as
experimentally observed membrane heights are close to the
self-assembled monolayer \cite{AVW:ASCN13}. Finally, the
system is cooled using thermostat dynamics. We test multiple
proportionality factors 
$k$ for the force profile ranging from
$30~\frac{\text{eV}}{\angstrom}$ to
$200~\frac{\text{eV}}{\angstrom}$
($z$ and $z_{\text{lo}}$ being dimensionless), which is
equivalent to a 
velocity range of $2.41~\frac{\angstrom}{\text{ps}}$ to
$16.07~\frac{\angstrom}{\text{ps}}$. 
Additionally, the same randomization as in the previous section
is applied to introduce some areas where bond formation might be
preferred. 

\begin{figure}[ht!]
\centering
\includegraphics*[clip,width=65mm,keepaspectratio]{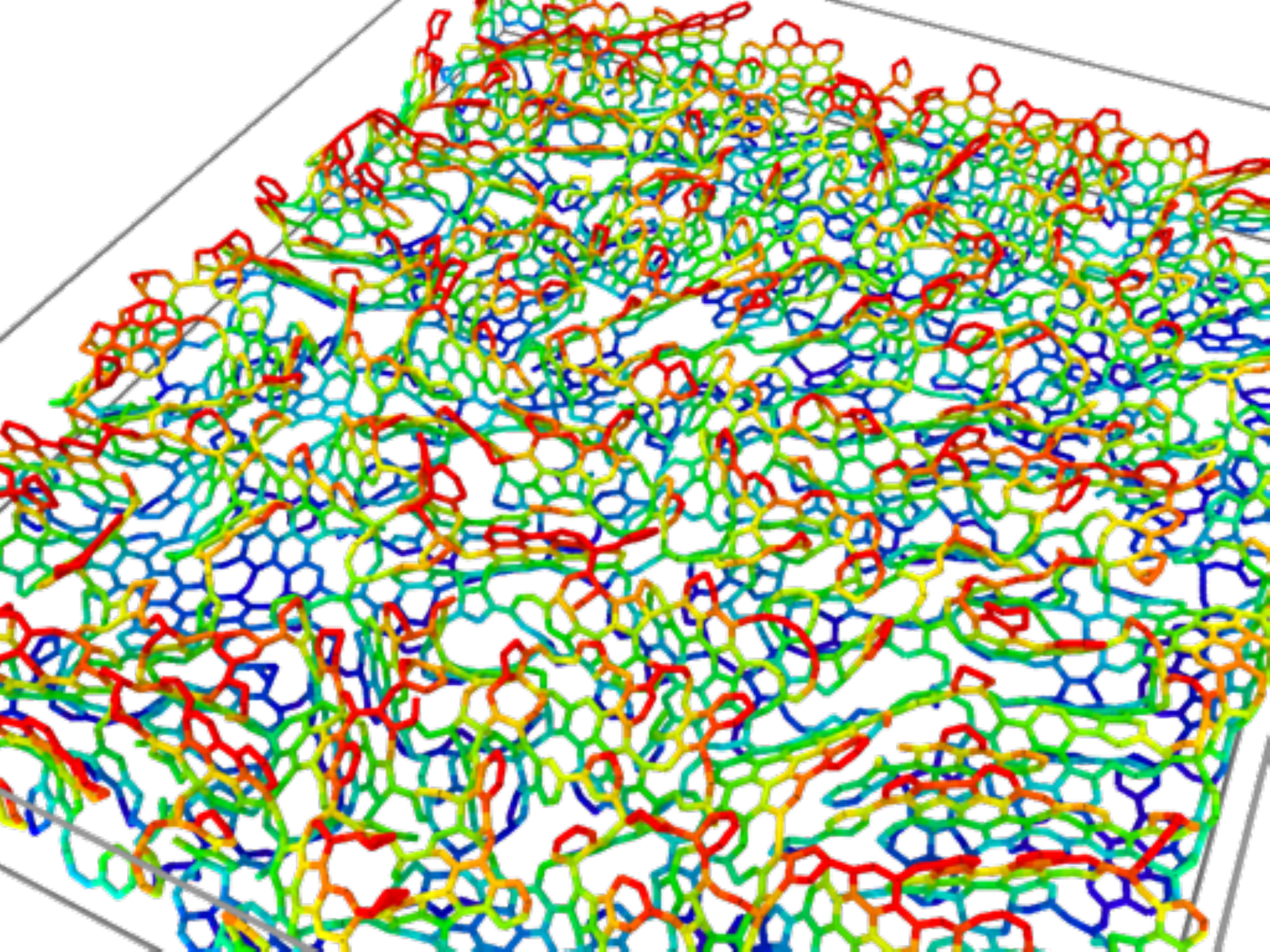}
\caption{A typical result of vertical momentum dynamics applied
  to TPT-SAM, $T=700$~K, $k=30~\frac{\text{eV}}{\angstrom}$. Color represents position along
  $z$-direction (blue to red as on regular maps).
  $x$- and $y$-directions as in \figref{wr-b}.}  
\label{v-t}
\end{figure}

\begin{figure}[ht!]
\centering
\includegraphics*[clip,width=65mm,keepaspectratio]{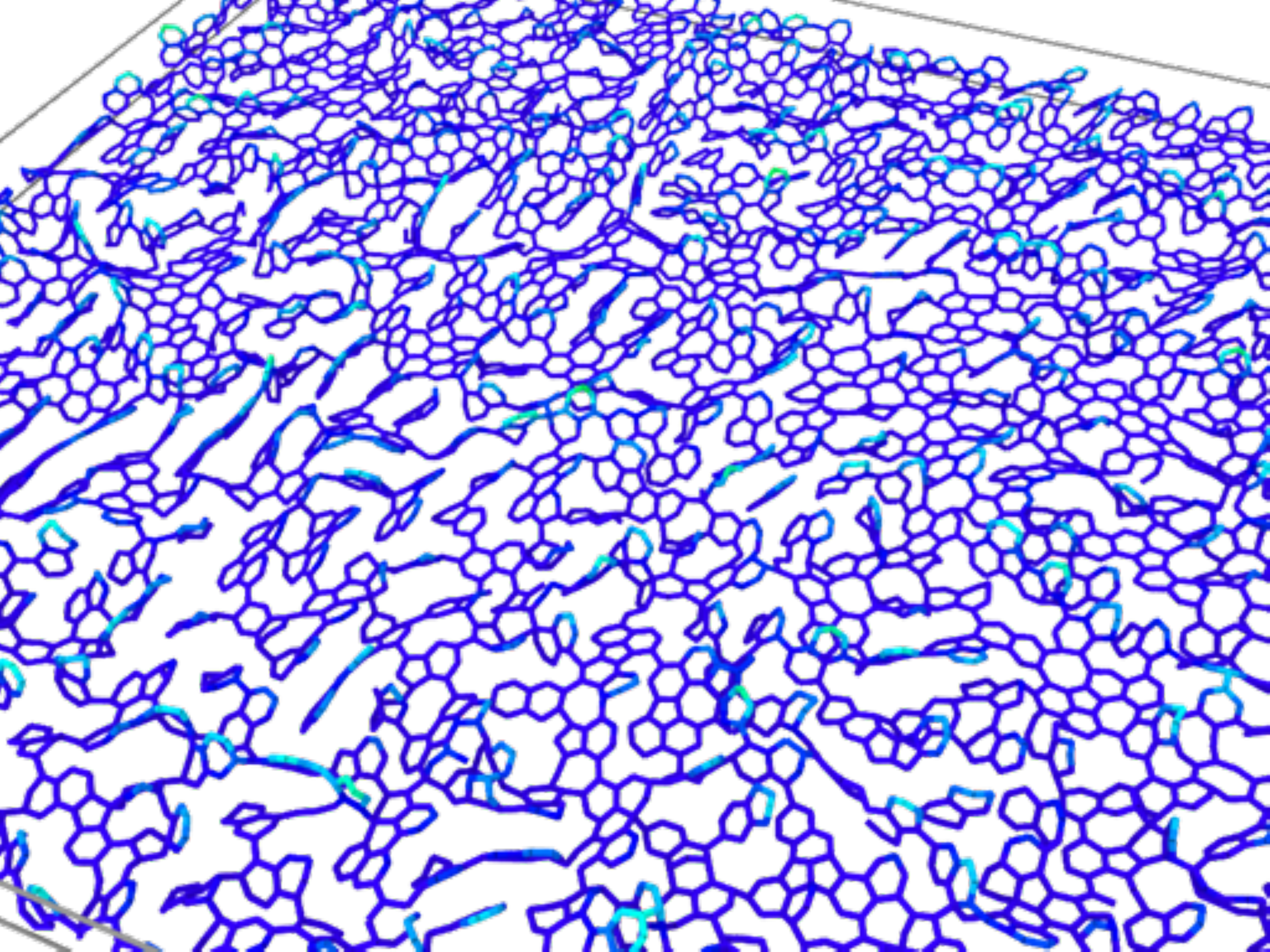}
\caption{A typical result of vertical momentum dynamics applied
  to NPTH-SAM, $T=700$~K, $k=30~\frac{\text{eV}}{\angstrom}$.  Color represents position along
  $z$-direction (blue to red as on regular maps). $x$- and $y$-directions as in \figref{wr-b}.} 
\label{v-n}
\end{figure}

Visualizations of membranes created through this process can be
seen in \figref{v-b} to \figref{v-n}. We note that the
resulting carbon networks are more irregular and contain
remnants of broken aromatic rings that serve as linkers in the
network. In particular, \figref{v-b} shows a typical result for
BPT. Since the precursor BPT consists of two phenyls only and a
large fraction of these are broken up, the resulting membrane is
rather flat (compare color code) and appears to consist of
denser regions that are loosely connected by phenyl
strings. Pentagonal structures, also reported in
\cite{CAS:PCCP10}, are visible.

Figure~\xref{v-t} presents a simulation of a TPT-based
nanomembrane. This membrane is thicker, since TPT consists of
three phenyls. The structure appears to be more strongly
connected in the $z$-direction (height). Voids and linear carbon
strings are visible.

Naphthalene on the other hand is a rather small molecule,
therefore the resulting CNM is rather flat and seems to contain
deformed graphene-like parts, see \figref{v-n}. We speculate
that the edge-sharing structure of the two rings in naphthalene,
compare Table~\xref{tab:sams}, is rather stable and promotes
graphene-like patches. 

The generated CNMs, which are mechanically stable, are
characterized by rather large Young's moduli, compare
Tables~\xref{tab:vertical1} and \xref{tab:vertical2}. 
In particular, naphthalene precursors, that form rather flat and
rigid membranes as discussed above exhibit moduli close to that
of graphene. Both tables show a systematic dependence on the
strength of the vertical force field characterized by $k$: The
larger $k$, the smaller is the modulus. Due to larger effective
randomization values along different directions are similar.
Interestingly, there is a systematic difference in moduli
between the two ways we compute them. Moduli calculated using the
AIREBO potential and applying stress are systematically smaller
compared to those using EDIP and the curvature of the potential
energy. Reasons need to be investigated in the future.

\begin{table}[ht!]
	\caption{Method 1 (EDIP and curvature): Young's moduli
          in $x$- and $y$-direction adjusted to volume of
          simulation box$|$surface volume for single
          realizations of membranes. 
          For direction of coordinates compare
          \figref{wr-b}.}
	\centering
	\begin{tabular}{@{}l|c|c@{}}
		\hline
		& $E_{x,1}$/GPa   & $E_{y,1}$/GPa       \\ \hline
		TPT (T=700~K, $k=30~\frac{\text{eV}}{\angstrom}$)& $436|847$ & $334|649$\\
		TPT (T=700~K, $k=200~\frac{\text{eV}}{\angstrom}$)& $215|448$ & $220|457$\\
		TPT (T=300~K, $k=60~\frac{\text{eV}}{\angstrom}$)& $325|987$ & $316|960$\\
		TPT (T=1100~K, $k=60~\frac{\text{eV}}{\angstrom}$)& $351|866$ & $339|838$\\
		BPT (T=700~K, $k=60~\frac{\text{eV}}{\angstrom}$)& $202|736$ & $191|695$\\
		NPTH (T=700~K, $k=60~\frac{\text{eV}}{\angstrom}$)& $536|1367$ & $500|1277$\\\hline
	\end{tabular}
	\label{tab:vertical1}
\end{table}

\begin{table}[ht!]
	\caption{Method 2 (AIREBO and stress): Young's moduli in
          $x$- and $y$-direction adjusted to volume of
          simulation box$|$surface volume for single 
          realizations of membranes. 
          For direction of coordinates compare
          \figref{wr-b}.}
	\centering
	\begin{tabular}{@{}l|c|c@{}}
		\hline
		& $E_{x,2}$/GPa   & $E_{y,2}$/GPa       \\ \hline
		TPT (T=700~K, $k=30~\frac{\text{eV}}{\angstrom}$)& $135|262$ & $77|150$\\
		TPT (T=700~K, $k=200~\frac{\text{eV}}{\angstrom}$)& $45|92$ & $40|83$\\
		TPT (T=300~K, $k=60~\frac{\text{eV}}{\angstrom}$)& $122|371$ & $97|295$\\
		TPT (T=1100~K, $k=60~\frac{\text{eV}}{\angstrom}$)& $123|303$ & $100|247$\\
		BPT (T=700~K, $k=60~\frac{\text{eV}}{\angstrom}$)& $16|58$ & $19|69$\\
		NPTH (T=700~K, $k=60~\frac{\text{eV}}{\angstrom}$)& $99|252$ & $45|115$\\\hline
	\end{tabular}
	\label{tab:vertical2}
\end{table}

\subsection{Additional lateral momentum}

In order to mimic the influence of secondary electrons and their
interaction with neighboring molecules and atoms, 
we incorporate additional lateral momenta of various magnitude
as shown in Tables \ref{tab:lateral1} and \ref{tab:lateral2}. We
apply an 
isotropic but randomly chosen lateral force to all atoms using the same LAMMPS fix
\verb|addforce| as before. Tables
\xref{tab:lateral1} and \xref{tab:lateral2} show averages over
10 realizations of such membranes depending on the theoretical
synthesis procedure.  By applying lateral momenta there is
a higher chance for holes to form due to displacement in the $x$-
and $y$-directions. This also affects membrane thickness and
surface roughness. This method relies highly on the randomly
chosen lateral force by which molecules are laterally
displaced. Realistically, forces would not be isotropic
throughout the membrane. Thus the Young's modulus is averaged
over ten different configurations each. Again, the values of
method~2 are systematically smaller compared to method~1 and
also closer to the experimentally determined moduli.

\begin{table}[ht!]
	\centering
	\caption{Method 1 (EDIP and curvature): Young's moduli
          in $x$-direction adjusted to volume of simulation
          box$|$surface volume. Numbers in parenthesis provide
          standard deviations for the averages taken over 10 realizations.}
	\begin{tabular}{@{}r|c|c|c@{}}
		\hline
		$v$ / $\frac{\angstrom}{\text{ps}}$ &  TPT: $E_{x,1}$/GPa & BPT: $E_{x,1}$/GPa   & NPTH: $E_{x,1}$/GPa \\ \hline
		$5$  & $338(55)|925(18)$ & $246(14)|782(12)$ & $588(41)|2002(47)$ \\
		$15$ & $299(20)|858(24)$ & $195(25)|888(15)$ & $546(32)|1865(38)$ \\
		$25$ & $224(46)|769(20)$ & $166(16)|818(24)$ & $497(25)|1579(49)$ \\
		$35$ & $268(46)|738(32)$ & $139(12)|732(12)$ & $410(34)|1393(40)$ \\ \hline
	\end{tabular}
	\label{tab:lateral1}
\end{table}

\begin{table}[ht!]
	\centering
	\caption{Method 2 (AIREBO and stress): Young's moduli in
          $x$-direction adjusted to volume of simulation
          box$|$surface volume. Numbers in parenthesis provide
          standard deviations for the averages taken over 10 realizations.}
	\begin{tabular}{@{}r|c|c|c@{}}
		\hline
		$v$ / $\frac{\angstrom}{\text{ps}}$ &  TPT: $E_{x,2}$/GPa & BPT: $E_{x,2}$/GPa   & NPTH: $E_{x,2}$/GPa \\ \hline
		$5$  & $120(23)|328(63)$ & $25(4)|79(13)$ & $201(32)|684(109)$ \\
		$15$ & $98(12)|281(34)$ & $23(5)|105(23)$ & $144(28)|492(96)$ \\
		$25$ & $62(14)|213(48)$ & $19(4)|94(20)$  & $114(34)|362(108)$ \\
		$35$ & $68(11)|187(30)$ & $16(3)|84(16)$ & $95(23)|323(78)$ \\ \hline
	\end{tabular}
	\label{tab:lateral2}
\end{table}

\begin{figure}[ht!]
\centering
\includegraphics*[clip,width=65mm,keepaspectratio]{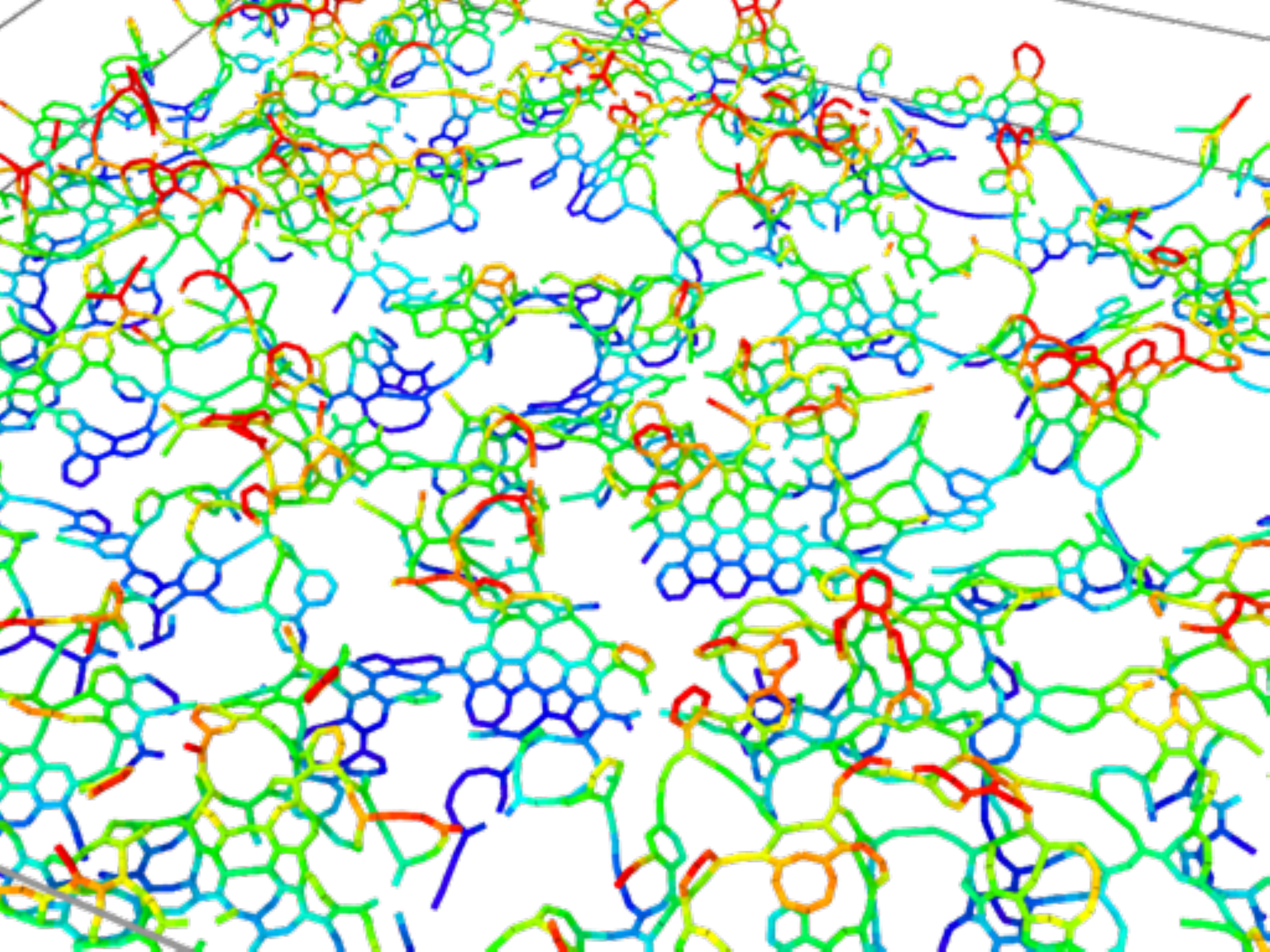}
\caption{A typical result of vertical and lateral momentum
  dynamics applied to BPT-SAM after 4900 time steps, $T=300$~K,
  $v=35~\frac{\angstrom}{\text{ps}}$,
  $k=60~\frac{\text{eV}}{\angstrom}$. Color represents position along
  $z$-direction (blue to red as on regular maps). $x$- and $y$-directions as in \figref{wr-b}.}  
\label{l-b}
\end{figure}

\begin{figure}[ht!]
\centering
\includegraphics*[clip,width=65mm,keepaspectratio]{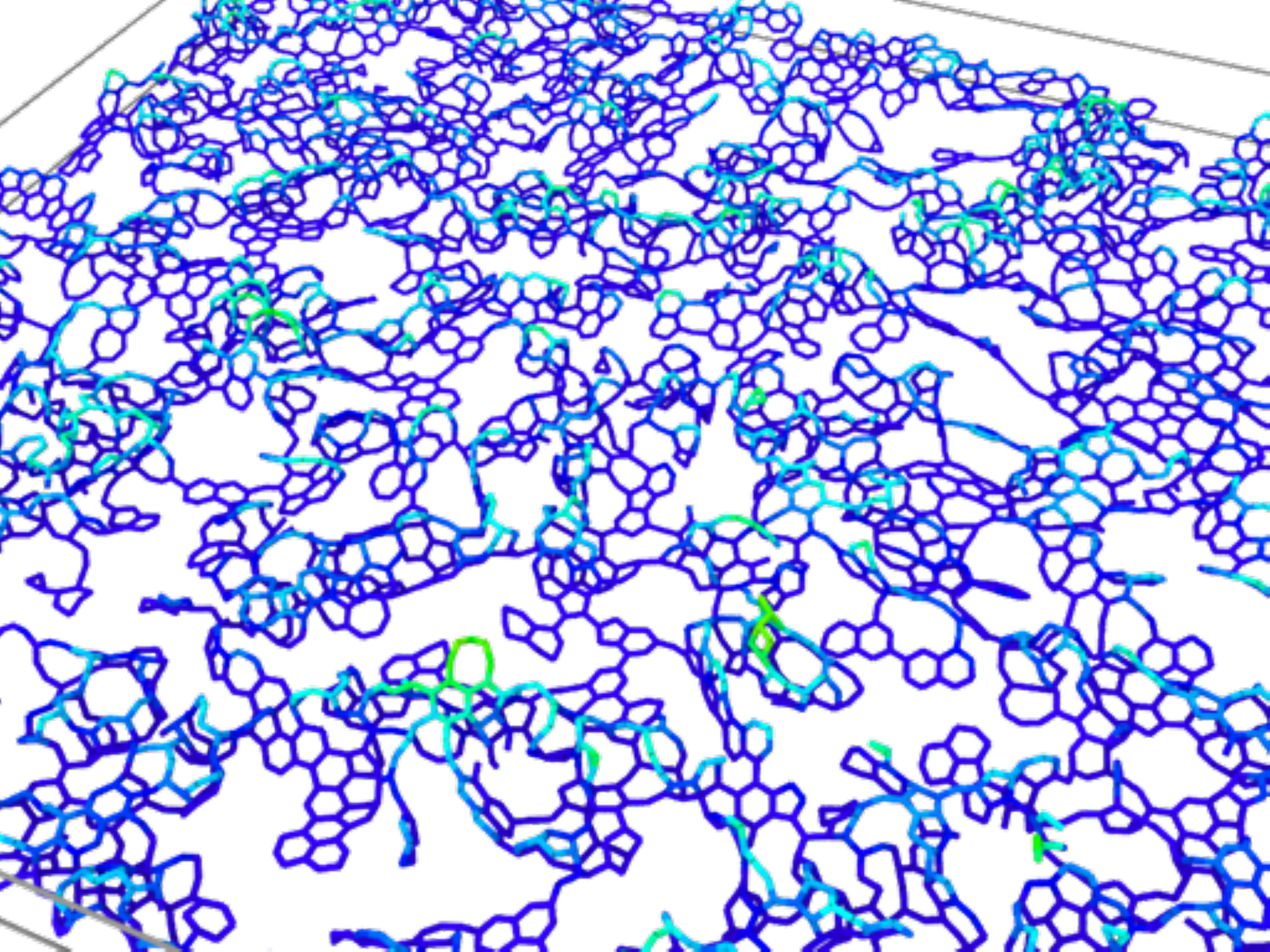}
\caption{A typical result of vertical and lateral momentum
  dynamics applied to NPTH-SAM after 2500 time steps, $T=300$~K,
  $v=35~\frac{\angstrom}{\text{ps}}$,
  $k=60~\frac{\text{eV}}{\angstrom}$. Color represents position along
  $z$-direction (blue to red as on regular maps). $x$- and $y$-directions as in \figref{wr-b}.}  
\label{l-n}
\end{figure}

\begin{figure}[ht!]
\centering
\includegraphics*[clip,width=65mm,keepaspectratio]{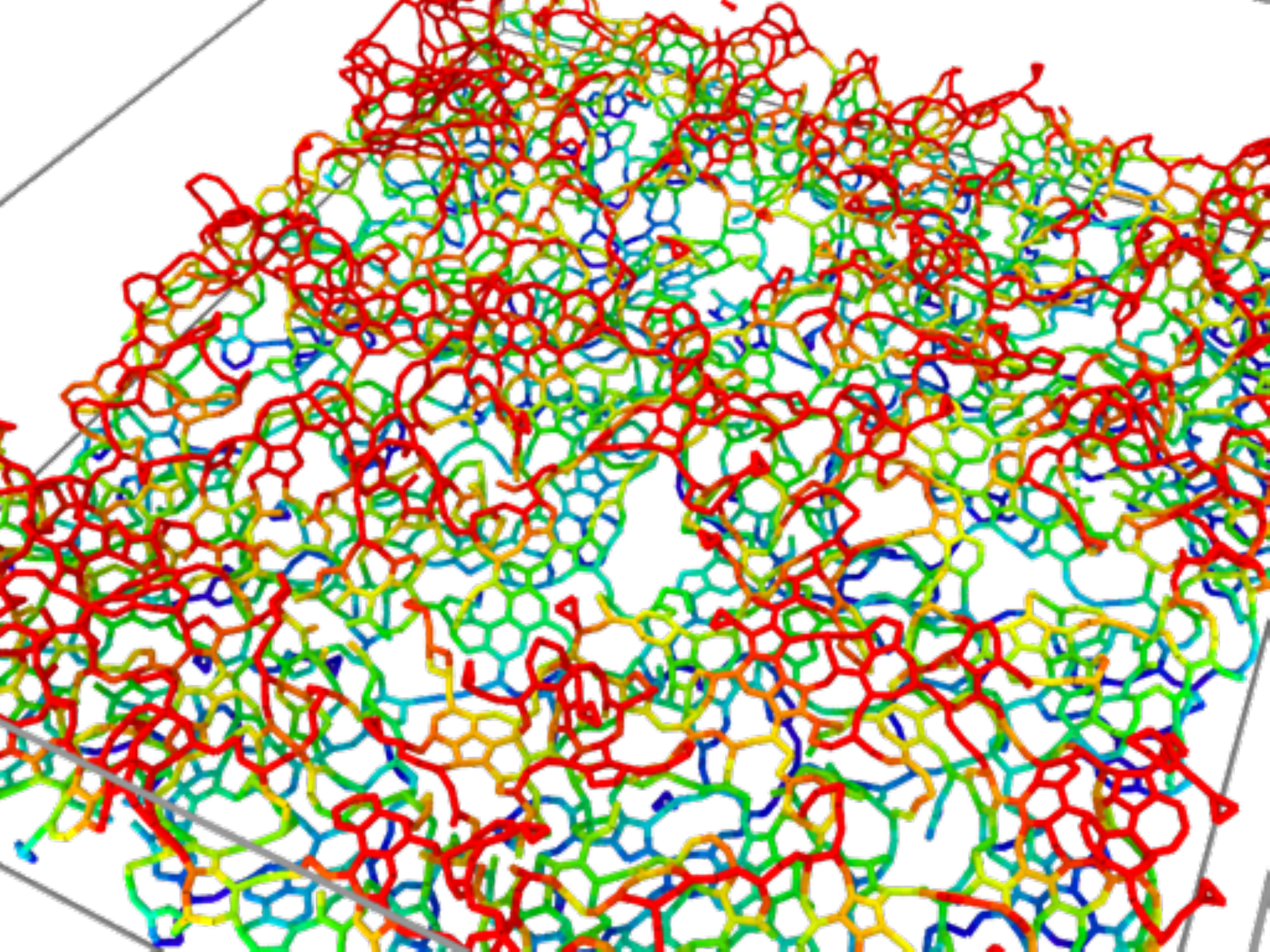}
\caption{A typical result of vertical and lateral momentum
  dynamics applied to TPT-SAM after 5700 time steps, $T=700$~K,
  $v=35~\frac{\angstrom}{\text{ps}}$,
  $k=30~\frac{\text{eV}}{\angstrom}$. Color represents position along
  $z$-direction (blue to red as on regular maps). $x$- and $y$-directions as in \figref{wr-b}.}  
\label{l-t}
\end{figure}

One can clearly see large holes for the  biphenyl-based and naphthalene-based CNMs in
Figs.~\ref{l-b} and \ref{l-n}, respectively. Both structures
appear to be bunched, i.e.\ they possess denser areas that are
connected by strings of carbon atoms or of phenyl rings.
For the realization of a terphenyl-based membrane depicted in
\figref{l-t}, holes appear not as pronounced, but there is
increased roughness compared to the previous
results, i.e.\ more pronounced valleys and hills in both height
and lateral distance from each other. Nevertheless, also TPT
nanomembranes exhibit an increased number of holes.

These qualitative results are also reflected in the quantitative
results for the Young's modulus, see Tables \xref{tab:lateral1}
and \xref{tab:lateral2}. With increasing magnitude of
the lateral force there is a decrease in the Young's modulus for
all precursor molecules. Only for the largest
$v=35\frac{\angstrom}{\text{ps}}$ the terphenyl-based
nanomembrane's Young's modulus increases, which might be related
to the height of the precursor molecule. While
biphenyl and naphthalene are basically two carbon rings tall,
terphenyl is about $50$~\% higher. This gives rise to the
possibility of bonds to form on top of the membrane allowing
increased surface roughness and more dense linking thereby
increasing the Young's modulus.

\subsection{Randomly missing molecules}

\begin{figure}[ht!]
\centering
\includegraphics*[clip,width=65mm,keepaspectratio]{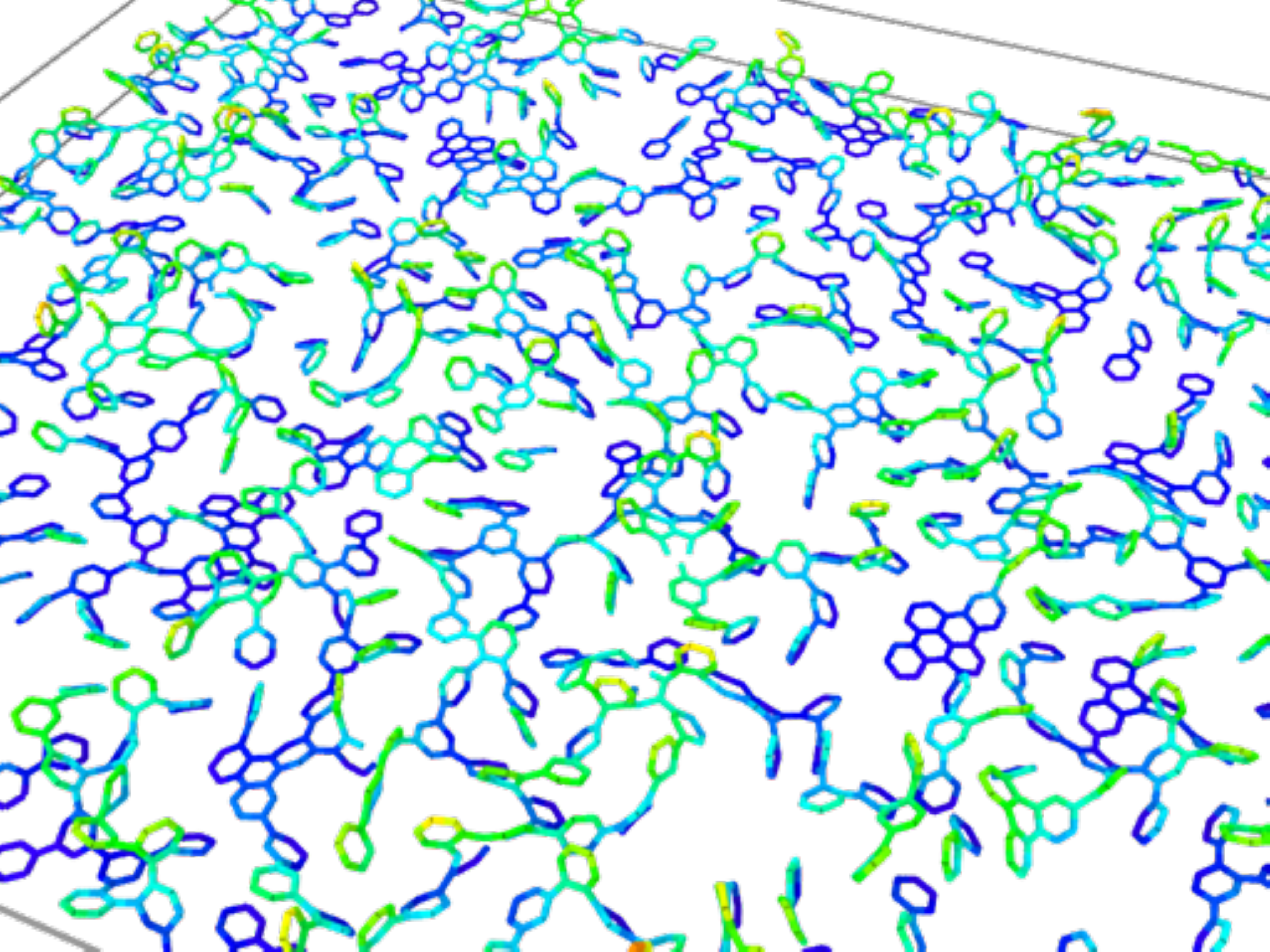}
\caption{A typical result of vertical momentum dynamics with
  missing molecules applied to BPT-SAM after 4900 time steps,
  $T=300$~K,  $k=60~\frac{\text{eV}}{\angstrom}$. Color represents position along
  $z$-direction (blue to red as on regular maps). $x$- and $y$-directions as in \figref{wr-b}.}  
\label{d-b}
\end{figure}

\begin{figure}[ht!]
\centering
\includegraphics*[clip,width=65mm,keepaspectratio]{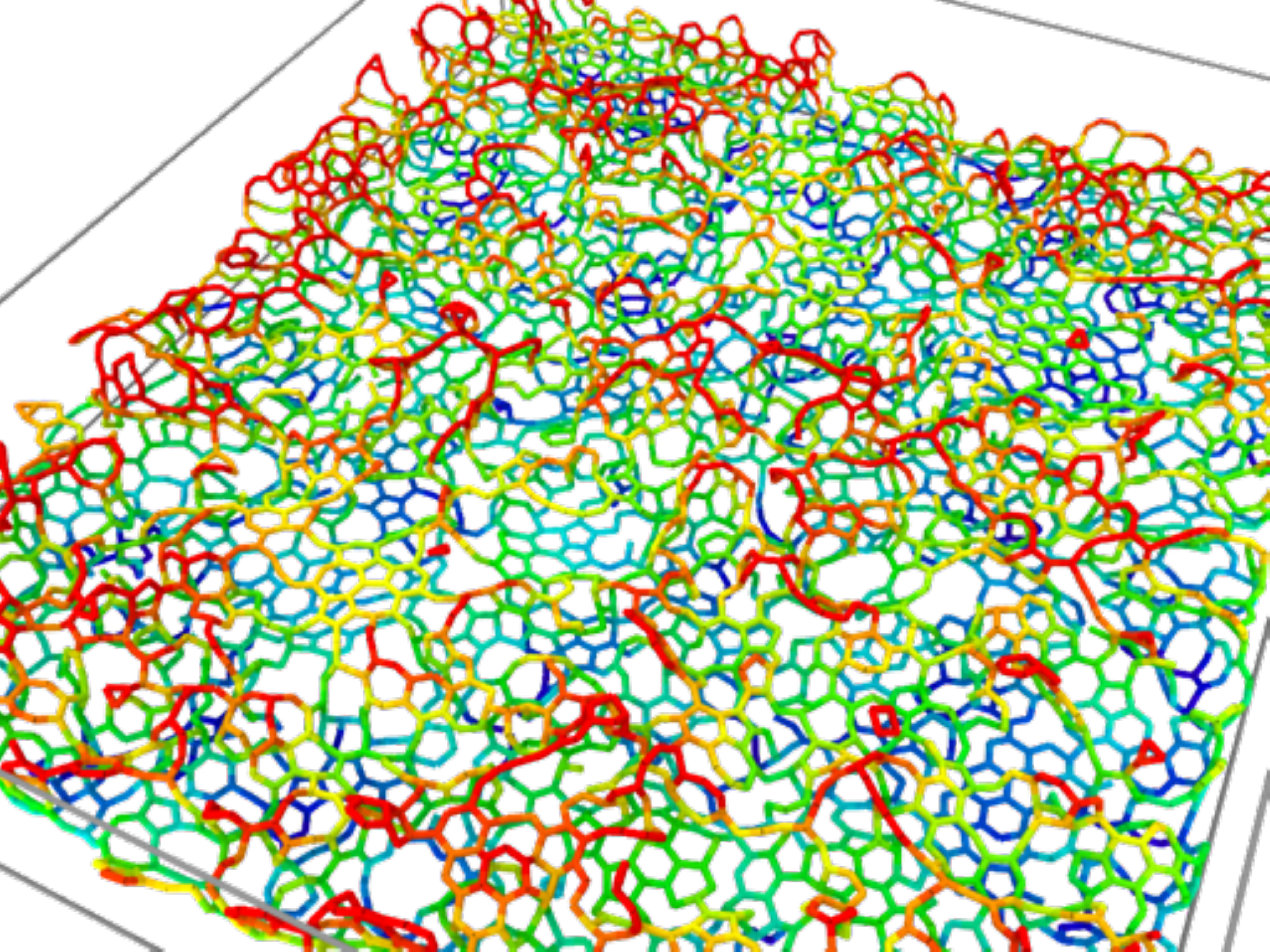}
\caption{A typical result of vertical momentum dynamics with
  missing molecules applied to TPT-SAM after 7200 time steps,
  $T=300$~K,  $k=60~\frac{\text{eV}}{\angstrom}$. Color represents position along
  $z$-direction (blue to red as on regular maps). $x$- and $y$-directions as in \figref{wr-b}.}  
\label{d-t}
\end{figure}

\begin{figure}[ht!]
\centering
\includegraphics*[clip,width=65mm,keepaspectratio]{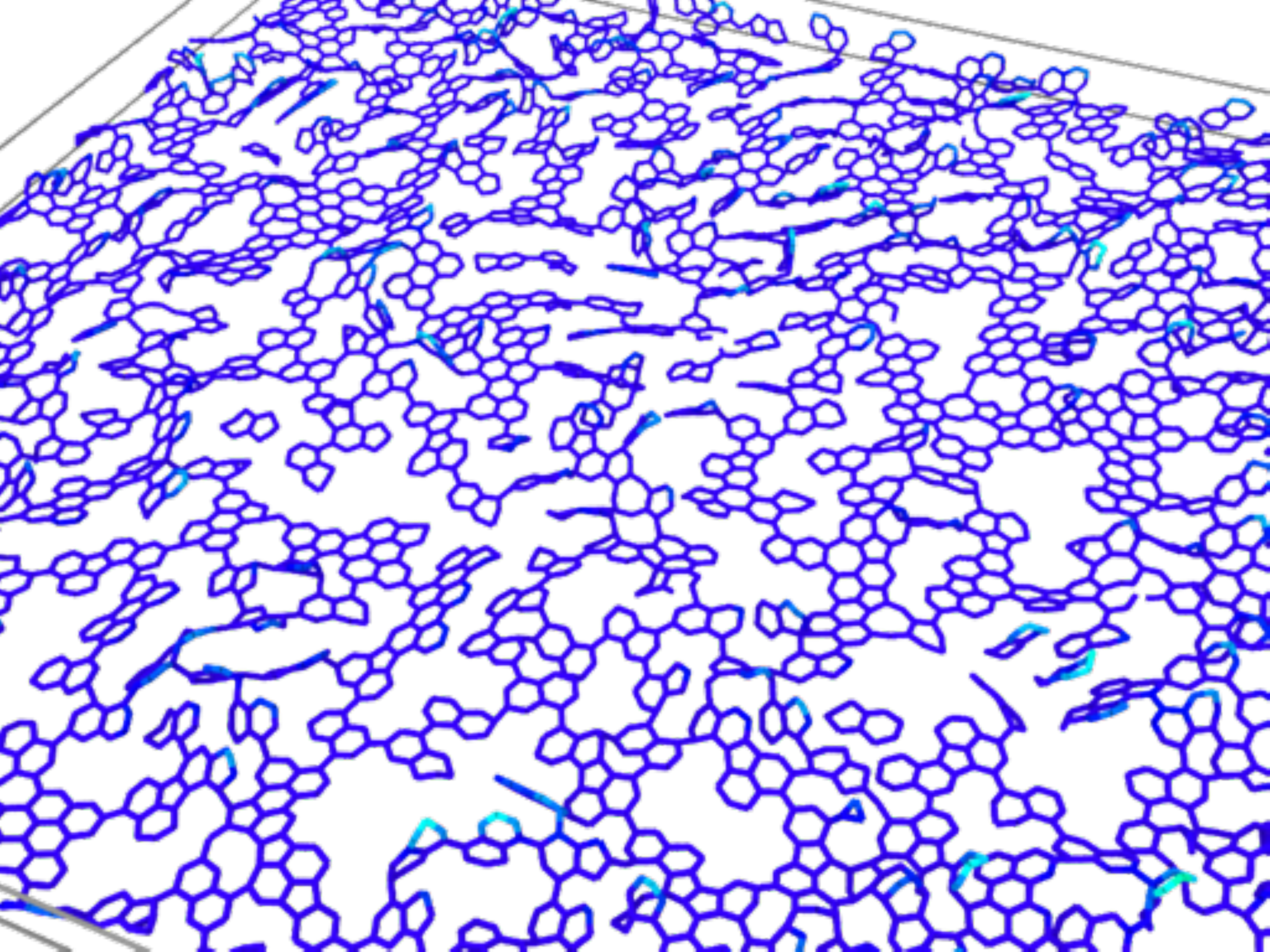}
\caption{A typical result of vertical momentum dynamics with
  missing molecules applied to NPTH-SAM after 2500 time steps,
  $T=300$~K,  $k=60~\frac{\text{eV}}{\angstrom}$. Color represents position along
  $z$-direction (blue to red as on regular maps). $x$- and $y$-directions as in \figref{wr-b}.} 
\label{d-n}
\end{figure}

By randomly removing molecules from the self-assembled monolayer
one can enhance the formation of holes in the resulting
membrane. It is experimentally verified that about 5 to 9~\%
of carbon atoms get lost during synthesis \cite{AVW:ASCN13}. Our
process models a correlated/collective disappearance of atoms in
form of whole molecules. We consider
percentages of removal ranging from $5$ to $20$~\%. 
Areas where molecules are missing are preferred
locations of holes as applied vertical momentum can only cover
the gaps to a limited degree. This also gives rise to the
possibility of further lowering the Young's modulus. The
resulting CNMs show a less dense structure than
before. Holes have the tendency to be smaller but more frequent
due to the more isotropic distribution of missing molecules,
which can be seen in Figs.~\xref{d-b}, \xref{d-t}, and
\xref{d-n}.

Again, BPT- and NPT-based membranes,
Figs.~\xref{d-b}, and \xref{d-n} turn out to be rather thin with
graphene-like patches. This seems to be a common and rather
stable motif under many conditions of preparation. The chosen
example of a TPT-based nanomembrane, \figref{d-t} appears rather
dense, even denser than the sample shown in \figref{l-t} that
had not experienced any random losses of precursor molecules.
This might be a coincidence and therefore calls for future large
scale simulations to generate sufficient statistics.

When it comes to quantitative results, the differences in Young's
moduli are not as pronounced as the visual differences, compare
Tables~\xref{tab:delete1} and \xref{tab:delete2}.
The moduli decrease by $10$ to $20$~\% at most. Even if there is
significant carbon loss when irradiating the SAM, the newly created
bonds are too isotropic to allow for softer areas. Thus any
local weak spot is corrected by molecules arranging flatter than
before. This is best observed for the naphthalene-based carbon
nanomembrane shown in \figref{d-n}, where one can see large
areas of intact hexagonal carbon rings strengthening the overall
membrane and yielding a larger modulus.

\begin{table}[ht!]
	\caption{Method 1: Young's moduli in $x$-direction for different deletion percentages adjusted to volume of simulation box$|$surface volume.}
	\centering
	\begin{tabular}{@{}r|c|c|c@{}}
		\hline
		$p$ / \% & TPT: $E_{x,1}$/GPa & BPT: $E_{x,1}$/GPa   & NPTH: $E_{x,1}$/GPa \\ \hline
		$5$  & $368|1011$  & $220|704$ & $620|2313$\\
		$10$ & $255|976$  & $177|787$ & $579|1689$\\
		$20$ & $329|1000$ & $193|824$ & $558|2437$\\ \hline
	\end{tabular}
	\label{tab:delete1}
\end{table}

\begin{table}[ht!]
	\caption{Method 2: Young's moduli in $x$-direction for different deletion percentages adjusted to volume of simulation box$|$surface volume.}
	\centering
	\begin{tabular}{@{}r|c|c|c@{}}
		\hline
		$p$ / \% & TPT: $E_{x,2}$/GPa & BPT: $E_{x,2}$/GPa   & NPTH: $E_{x,2}$/GPa \\ \hline
		$5$  & $131|360$  & $21|67$ & $135|503$\\
		$10$ & $86|329$  & $16|71$ & $50|146$\\
		$20$ & $105|319$ & $16|68$ & $40|175$\\ \hline
	\end{tabular}
	\label{tab:delete2}
\end{table}

\section{Discussion and Conclusions}
\label{sec-5}

Our goal was to create first computer simulations that model the
process of CNM formation as realistically as possible. In order
to achieve this goal we suggest various scenarios abstracting
the experimental process such that the formation can be modeled
by classical molecular dynamics.

We have shown that some processes deliver membranes that appear
closer to the experiment than others. The most violent
approaches, applying both vertical and lateral momentum
transfer, are able to produce better results with respect to the
visual impression of the 
membrane in particular concerning holes needed for its
filtration abilities. This is a crucial step in 
understanding the internal structure of the membrane and
possible molecular and atomic processes involved.

Results fall shorter when it comes to reproducing the experimental value
of the Young's modulus, which experimentally can be determined
by various means, e.g.\ bulge tests, to be at least an order of magnitude
smaller than our results. This is where the layers of
abstraction play a big role. There are no hydrogen atoms and
electrons present in our model system. Thus, breaking
carbon-hydrogen bonds and momentum transfer by hydrogen atoms is
only effectively included. Missing electron dynamics may
strongly simplify the intricate momentum transfer, e.g.\ by
secondary electrons. 

However, all abstractions have to be made in order to be able to
simulate large enough systems of carbon atoms and thus a
reasonably sized area of a membrane. Other more accurate
simulations, e.g.\ done by density functional theory (DFT) are
limited to rather small numbers of atoms of order $10^2$ \cite{MCM:PRB02} or
have to introduce periodicity into the simulation
\cite{CAS:PCCP10}, which is a clear bias.

We are nevertheless confident that further progress will be
possible. For example, in future investigations selected membranes
will be simulated that contain a certain amount of hydrogen. This
requires the use of classical carbon-hydrogen potentials such as
for instance AIREBO. On the experimental side, it is planned to
do ion scattering in order to determine structural correlations
of CNMs \cite{WiG:CP19,Wil.JPCS20,Wil:PC21}. 

A problem inherent to all simulations of atomic processes
consists in the large span of time scales. The intrinsic time
step is of order picosecond whereas some processes might need
milliseconds. This means that one usually cannot model the
relatively slow relaxation into the final state. Here
workarounds have been developed recently \cite{TSV:C17} that
artificially
speed up such processes by an appropriately chosen higher
temperature. It will also be investigated whether such a
treatment would modify the theoretical formation process.

In a future investigation we are going to start a mass
production of simulations for better statistics, in particular
hole distributions as well as size distributions in dependence
of production conditions.
This is also needed for the idea to compare experimental AFM
images with surfaces of simulated membranes. We would like to
show first attempts along these lines in the appendix.

\section*{Acknowledgment}
We are very thankful to Nigel Marks for sharing with us
the details of his EDIP carbon-carbon potential.


%

\appendix

\section{Stylized model (artist's view) of atomic force microscopy}
\label{sec-a-1}

One approach to create a visual impression of a
nanomembrane is to employ atomic force microscopy (AFM) as e.g.\ done
in \cite{yang2018rapid}. The basic procedure is to rasterize the
membrane using a cantilever and by that render atomic structures
as well as valleys and holes visible. However, due to limited
resolution of this approach, there is no certainty as to
whether there are real holes or channels in the membrane. As
demonstrated above, the visualization of
molecular dynamics simulations is perfect in the sense that it
shows the actual position of atoms and bonds. It is thus hard 
to compare these results to experimental atomic force microscopy
images.

\begin{figure}[ht!]
\centering
\includegraphics*[clip,width=75mm,keepaspectratio]{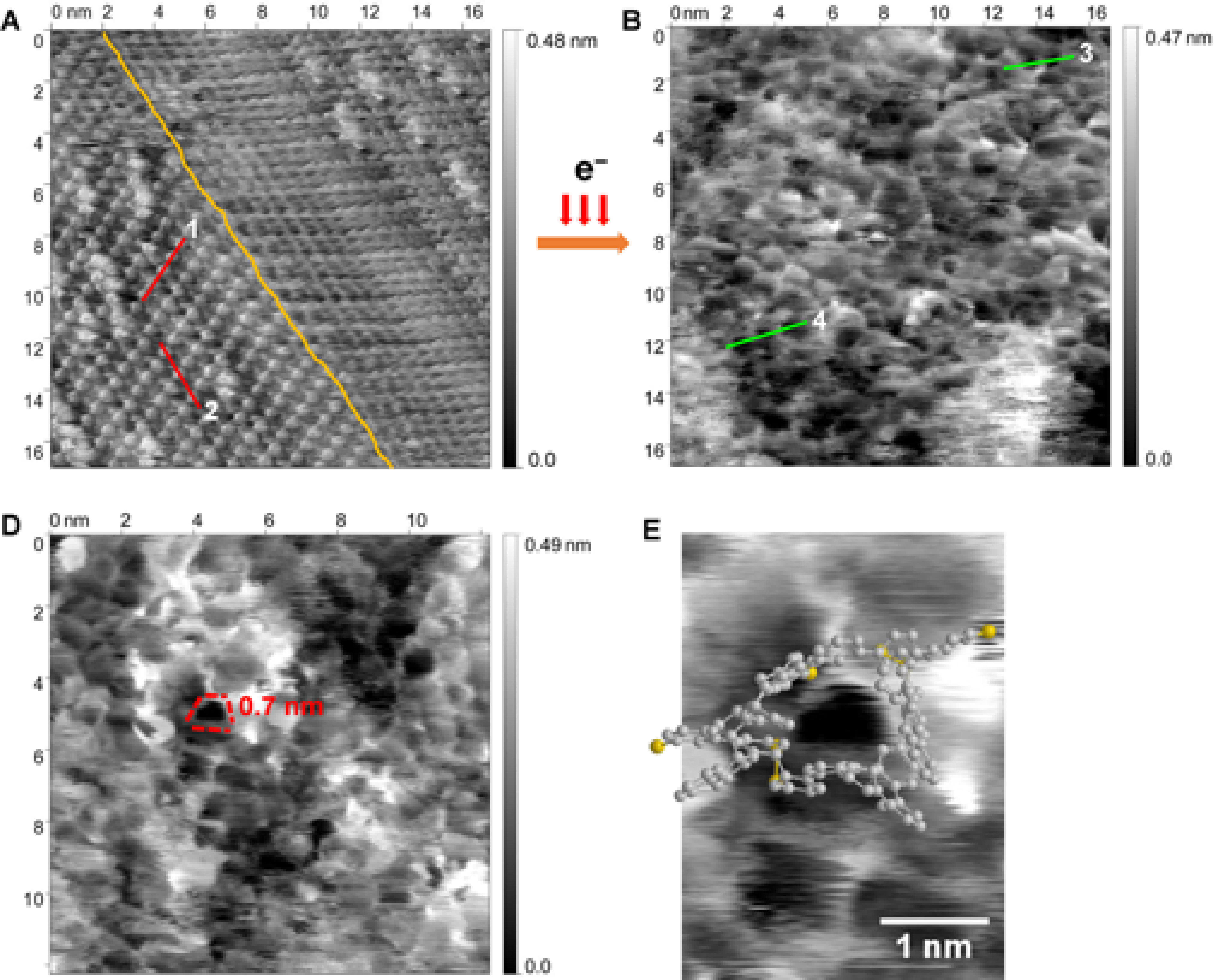}
\caption{Experimental scanning probe and atomic force microscopy
  images of a TPT-based CNM \cite{yang2018rapid} (with 
  friendly permissions). See also Ref.~\cite{BKS:JPCC19}.} 
\label{afm-real}
\end{figure}

\begin{figure}[ht!]
\centering
\includegraphics*[clip,width=65mm,keepaspectratio]{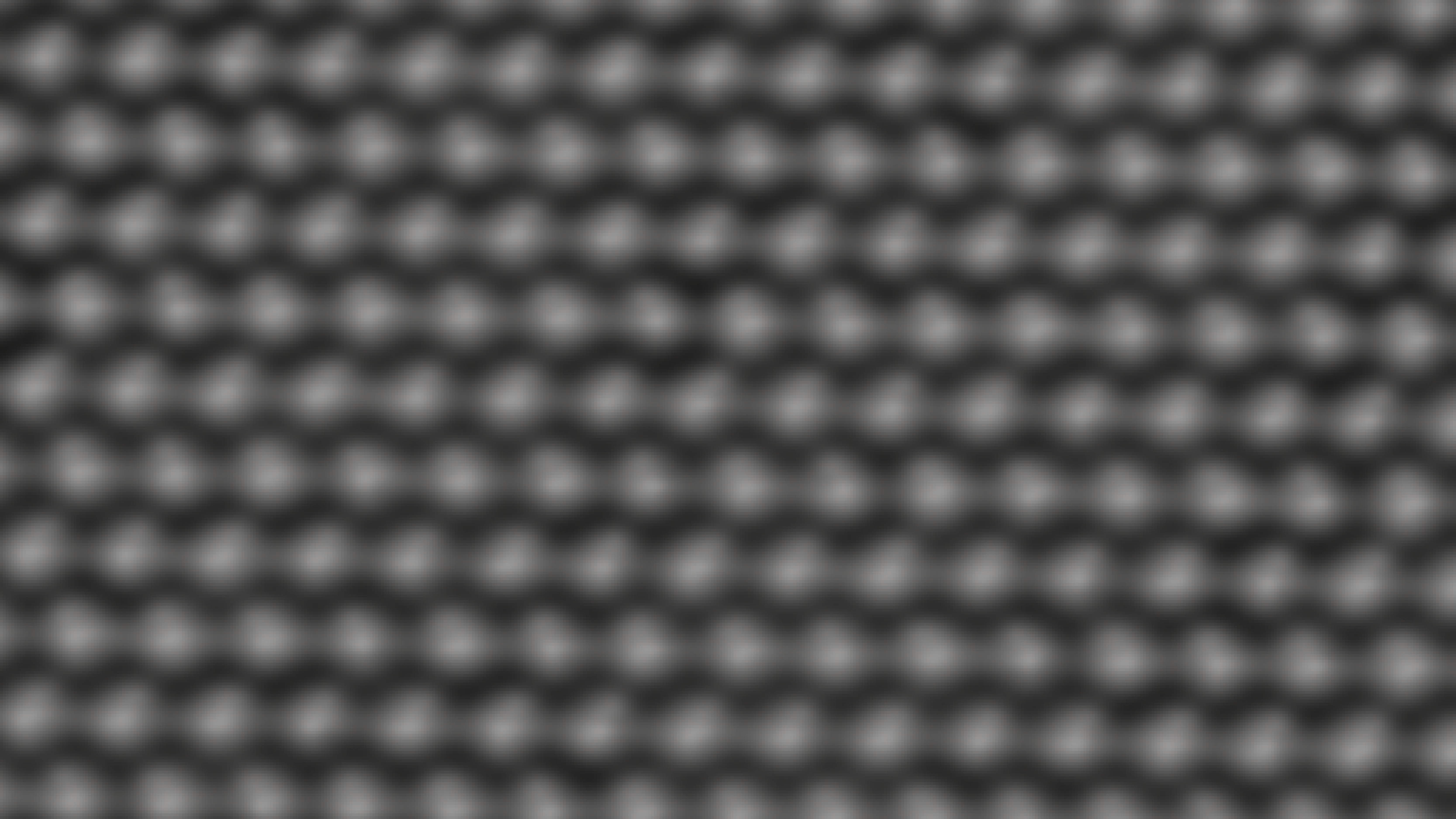}
\caption{Simulated atomic force microscopy image of a
  TPT-based self-assembled monolayer.} 
\label{afm-ter-precursor}
\end{figure}

\begin{figure}[ht!]
\centering
\includegraphics*[clip,width=65mm,keepaspectratio]{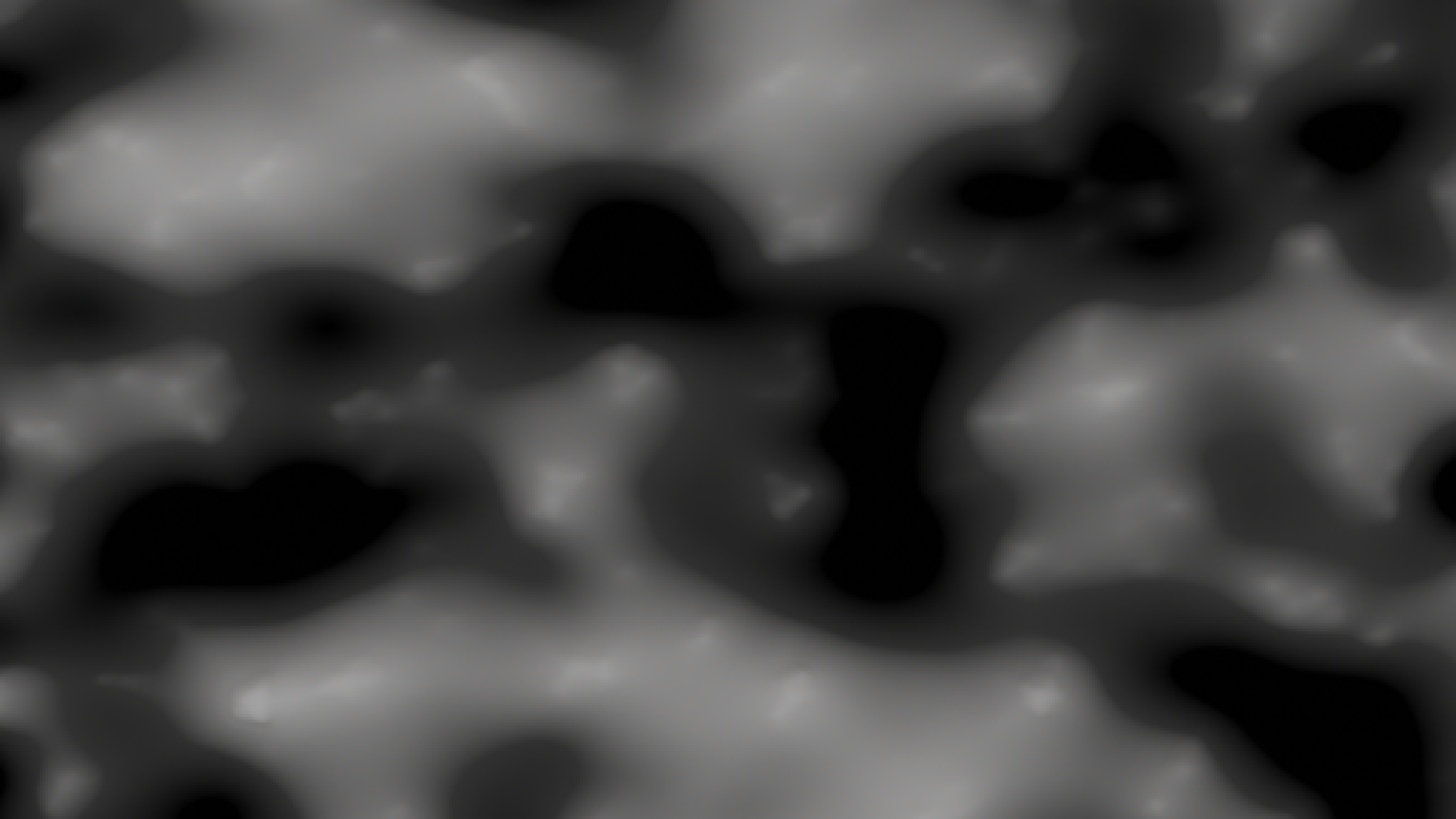}

\vspace*{1mm}

\includegraphics*[clip,width=65mm,keepaspectratio]{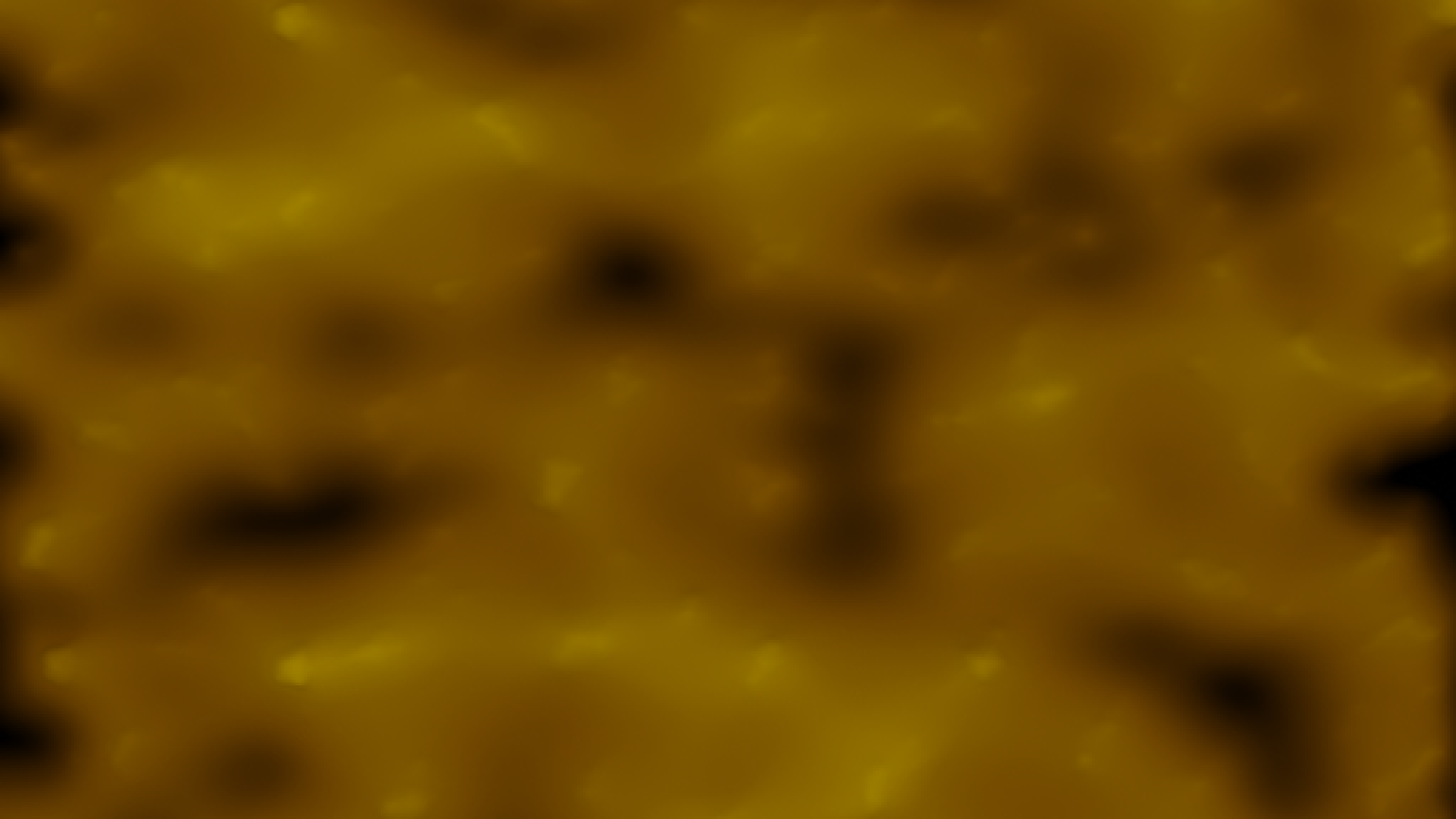}
\caption{Two versions of a simulated atomic force microscopy image of the
  TPT-based CNM shown in \figref{l-t}. The chosen theoretical
  color code determines the impression of depth strongly.} 
\label{afm-ter-membrane}
\end{figure}

Here we present a stylized and artistic approach to generate
images that have the same color scheme, i.e.\ representation of
height, as atomic force microscopy and have (artificially) limited resolution
as to which smallest structures can be resolved. It should be
noted that this is by no means a quantitative measure as it is
highly dependent on degrees of freedom of visualization
parameters. The images have been created using the open-source
software Blender \cite{blender}. 

An experimental result of atomic force microscopy of a TPT-based
membrane taken from Ref.~\cite{yang2018rapid} is shown in
\figref{afm-real}. Panel A shows the SAM, panel B the
crosslinked CNM; D and E display a possible hole and a
hypothetical arrangement of TPT around the hole.
For comparison,
\figref{afm-ter-precursor} shows the
initial self assembled monolayer of terphenyls as used in our
simulations, and \figref{afm-ter-membrane} displays a resulting
membrane, respectively. 
The parameters for this membrane are
$T=700$~K, $v=35~\frac{\angstrom}{\text{ps}}$ and
$k=30~\frac{\text{eV}}{\angstrom}$; the membrane is the same as
in \figref{l-t}. Figure \ref{afm-ter-membrane} also demonstrates that
the chosen theoretical color code determines the impression of
depth rather strongly. Nevertheless, this might be of great help
in interpreting similar experimental pictures in order to
unambiguously identify holes.


\section{Investigated precursor SAMs}\label{sams}

\onecolumngrid

\begin{table}[ht!]
	\centering
	\caption{Precursor molecules and structures of the
          respective self assembled monolayers.}
	\label{tab:sams}
	\begin{tabular}{p{4.1cm}|l|p{3.7cm}|l}
		\toprule
		Name & Structural formula & SAM structure &\\ \hline

		Biphenyl-4-thiol (BPT)\newline 1,1'-Biphenyl-4-thiol\newline 4-Biphenylylthiol\newline 4-Mercaptobiphenyl\newline 4-Phenylbenzenethiol&
		\includegraphics[scale=0.15,valign=t]{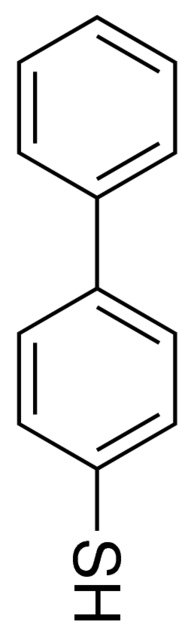}\cite{biphenyl}
		& (2 x 2) hexagonal, \newline$\gamma = 30^\circ$~\cite{structuralInvest} \newline
		$\gamma = 15^\circ$~\cite{GSE:APL99}\newline
		$\gamma = 20^\circ$~\cite{structureOfThioaromatic}\newline
		$\gamma = 20^\circ$~\cite{oddEvenEffects}&
\includegraphics*[scale=0.16,valign=t]{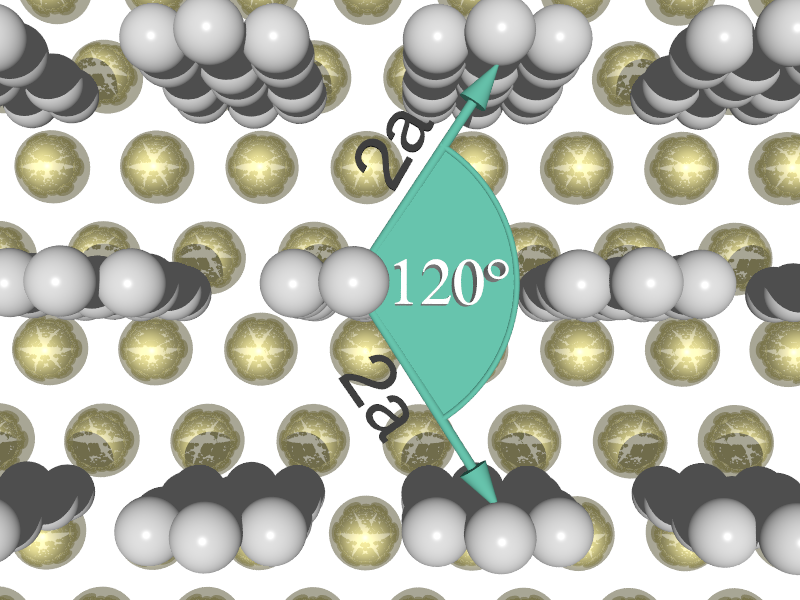}
		\\ 
		&& mixture of:\newline (2 x 2) structure,\newline (2$\sqrt{3}$ x 9) unit cell,\newline (2$\sqrt{3}$ x 8)
		unit cell~\cite{structuralInvest}
		\\\hline

		1,1',4',1''-Terphenyl-4-thiol\newline (TPT)&
\includegraphics[scale=0.2,valign=t]{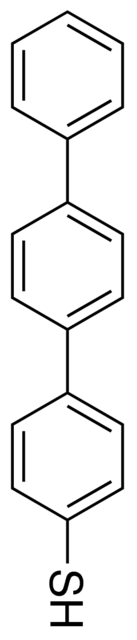}\cite{terphenyl}&
		$(\sqrt{3}$ x $\sqrt{3})$ structure \newline $(2\sqrt{3}$ x $\sqrt{3})$ unit cell\newline $\gamma = 20^\circ$~\cite{structureOfThioaromatic}
&\includegraphics[scale=0.16,valign=t]{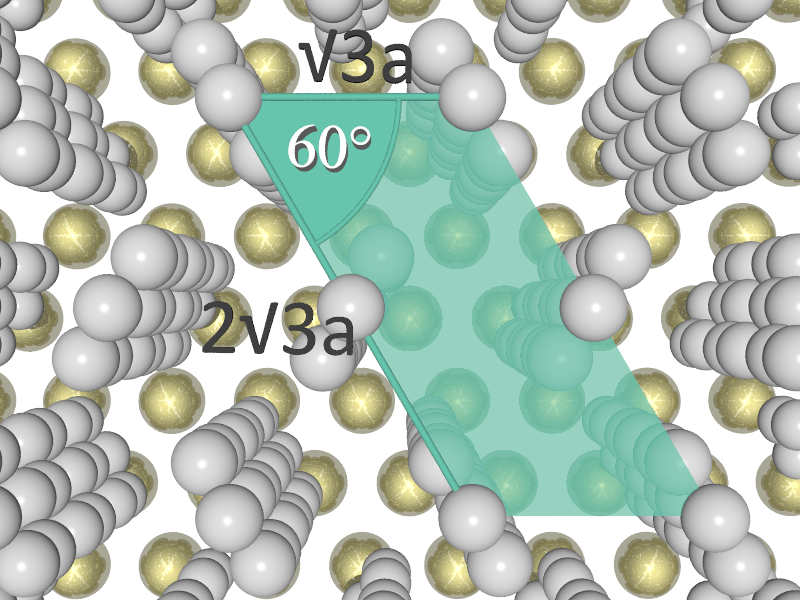}
                \\ \hline

		2-Naphthalenethiol (NPTH)\newline 2-Naphthyl mercaptan \newline Thio-2-naphthol ($\beta$)&
\includegraphics[scale=0.12,valign=t]{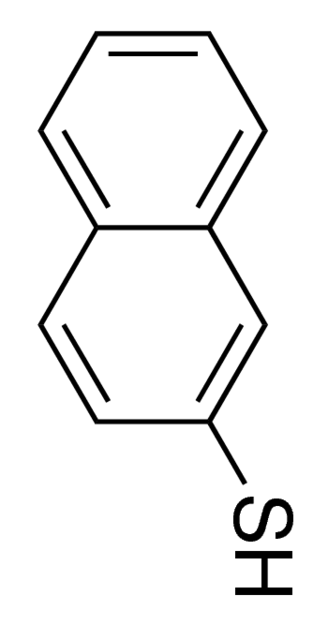}\cite{naphtalen}&
		$(\sqrt{3}$ x $\sqrt{3})$ structure \newline $(2\sqrt{3}$ x $\sqrt{3})$ unit cell
&\includegraphics[scale=0.16,valign=t]{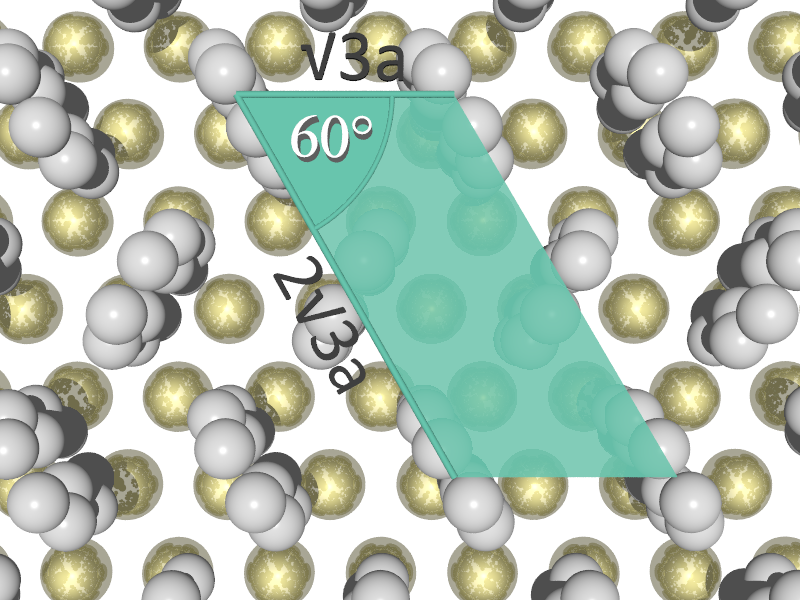}
                \\ \hline
	\end{tabular}
\end{table}

\twocolumngrid

\end{document}